\documentclass[a4paper,11pt]{article} 
\usepackage{amsmath} 
\usepackage{amsthm} 
\usepackage{amssymb}	
 \usepackage{indentfirst}
\usepackage{float}
\usepackage{titlesec}
\usepackage[T1]{fontenc}
\titlelabel{\thetitle. \quad}
\usepackage{authblk}
\usepackage{graphicx} 
\usepackage[dvips,letterpaper,margin=0.75in,bottom=0.5in]{geometry}
\usepackage[titletoc,toc,title]{appendix}
\usepackage{verbatim}
\usepackage{cite}
\usepackage{tikz}
\usetikzlibrary{tikzmark}
\usetikzlibrary{shapes,arrows}
\usepackage{mathtools,bbm}
\usepackage{amsfonts}
\usepackage[percent]{overpic}
\usepackage{cancel,url}
\usepackage{appendix}
\usepackage{color,xcolor}
\usepackage{tabularx, booktabs}
\definecolor{darkred}{rgb}{0.4,0.0,0.0}
\definecolor{darkgreen}{rgb}{0.0,0.4,0.0}
\definecolor{darkblue}{rgb}{0.0,0.0,0.4}
\usepackage[bookmarks,linktocpage,colorlinks,
    linkcolor = darkred,
    urlcolor  = darkblue,
    citecolor = darkgreen]{hyperref}
\numberwithin{equation}{section}
\newcolumntype{Y}{>{\centering\arraybackslash}X} 
\newcommand{\ben}{\begin{enumerate}}
\newcommand{\een}{\end{enumerate}}
\newcommand{\bit}{\begin{itemize}}
\newcommand{\eit}{\end{itemize}}
\newcommand{\bsa}{\begin{subequations}\begin{eqnarray}}
\newcommand{\esa}{\end{eqnarray}\end{subequations}}
\newcommand{\bean}{\begin{eqnarray*}}
\newcommand{\ean}{\end{eqnarray*}}

\newcommand{\req}{\overset{\footnotesize!}{=}}
\newcommand{\ist}[1]{\overset{\footnotesize(\ref{#1})}{=}}

\newcommand{\LR}[1]{\overset{\footnotesize(\ref{#1})}{\Leftrightarrow}}
\newcommand{\zu}[1]{\overset{\footnotesize(\ref{#1})}{\to}}

\begin{document}

\title{Chiral Symmetry Breaking on the Lattice}

\author[1]{Manfried Faber} 
\affil[1]{Nuclear Physics Department, Institute of Atomic and Subatomic Particles, Vienna University of Technology, Operngasse 9, 1040 Wien, Austria}

\author[2]{Roman H\"ollwieser}
\affil[2]{Department of Physics, Fakult\"at f\"ur Mathematik und Naturwissenschaften, 
Bergische Universit\"at Wuppertal, Gau{\ss}stra{\ss}e 20, 42119 Wuppertal, Germany}

\maketitle

\begin{abstract} 
We review important aspects of QCD in the continuum and on the lattice and take a look at the fate of
its symmetries with an emphasis on chiral symmetry breaking on the lattice.
\end{abstract}

\tableofcontents

\section{Introduction}
Like any relativistic quantum field theory, QCD enjoys Poincar\'e symmetry including the discrete symmetries charge conjugation $C$, parity $P$ and time reversal $T$, each of which is realized. Apart from these space-time symmetries, it also has internal symmetries. Since QCD is an $SU(3)$ gauge theory, it has local $SU(3)$ gauge or color symmetry. 

Due to their spin we can assign a handedness or helicity to fermions. Their fields can be decomposed into left- and right-handed components. The mass of fermions couples these components, it breaks chiral symmetry. For massless quarks the QCD-Lagrangian does not have an interaction term between the two quark chiralities, a coupling term like the Nambu-Jona-Lasinio model, breaking chiral symmetry explicitly. Due to the missing interaction term, left- and right-handed fermions can be transformed independently without modifying the Lagrangian.

Up to mass differences of a few percent hadrons can be grouped in multiplets with the same isosopin $I$ reflecting an approximate $SU(2)_I$ isospin symmetry of the Lagrangian. The corresponding isospin transformations act on left- and right-handed fermions simultaneously, {\textit i.e.}, $SU(2)_I=SU(2)_{L=R}$. The $SU(3)_F=SU(3)_{L=R}$ flavor symmetry is violated more strongly as seen in the octet of pseudo-scalar mesons. The axial vector symmetry $SU(N_f)_A=SU(N_f)_{L=R^\dagger}$ the chiral symmetry, on the other hand, is not manifest in the spectrum at all. The experimental evidence for the absence the $SU(N_f)_A$ symmetry is twofold. Axial transformations mix states with different parity. But in the low-lying hadron spectrum one does not observe the corresponding mass-degenerate parity doublets, states with the same quantum numbers, besides parity. The second indication is the above mentioned appearance of (nearly) massless Goldstone bosons. As soon as the chiral symmetry is dynamically broken at low momenta, then necessarily appear Goldstone bosons. We conclude that in massless QCD chiral symmetry is ``spontaneously'' or ``dynamically'' broken, it is realized in the Nambu-Goldstone mode, the Lagrangian is chiral symmetric but the vacuum is not. 

Despite the absence of interaction terms between left- and right-handed fermions, quarks and anti-quarks are bound in pions to spin-zero states of negative parity. Besides the very light pions $\pi^+$, $\pi^0$, and $\pi^-$ one observes somewhat heavier pseudo-scalars, the four kaons $K^+$, $K^-$, $K^0$, $\overline{K^0}$ and the $\eta$-meson. According to Goldstone's theorem, the number of massless bosons is given by the difference of the number of generators of the full symmetry group $G$ and the subgroup $H$ that remains unbroken. In massless QCD the full chiral symmetry group is
\begin{equation}
G = SU(N_f)_L \otimes SU(N_f)_R \otimes U(1)_B,
\end {equation}
while the unbroken subgroup is the flavor symmetry
\begin{equation}
H = SU(N_f)_{L=R} \otimes U(1)_B.
\end{equation}
Hence, in this case one expects $N_f^2 - 1$ massless Goldstone bosons. For $N_f=2$ there is the isovector triplet of pions with $m_\pi\approx140$~MeV indicating that in the groundstate of QCD the axial vector symmetry is broken, while for $N_f = 3$ there are eight Goldstone bosons --- the pions, the kaons, and the $\eta$-meson. In nature these particles are not exactly massless, because chiral symmetry is explicitly broken by the quark masses. The masses of the up and down quarks are much smaller than the QCD scale $\Lambda_{\overline{MS}}\approx300$~MeV ($N_f = 3$), which leads to the very small pion mass. The mass of the strange quark, on the other hand, is of the order of $\Lambda_{\overline{MS}}$, thus leading to larger masses of the kaons and the $\eta$-meson. Still, their masses are small enough to identify these particles as pseudo-Goldstone bosons. 
In the classical massless theory for $N_f$ massless flavors there would be an independent $U(N_f)$ symmetry associated with each chirality which can be combined to vector and axial vector symmetries $U(N_f)_V\times U(N_f)_A$, see Sect.~\ref{Sect:ContForm}. Sect.~\ref{Sect:AnomClass} explains why this full symmetry does not survive quantization, being broken to the above mentioned $SU(N_f)_V\times SU(N_f)_A\times U(1)_B$. For finite quark masses of these chiral symmetries, only the baryon number symmetry $U(1)_B$ is exact.

Due to the dimensionless coupling constant $g$ classical QCD is approximately scale invariant, $x_\mu\to\lambda x_\mu$, for small quark masses. Like $U(1)_A$ this classical symmetry is broken by quantum fluctuations, the scale symmetry is anomalous, \cite{Shuryak:1977ut,Landsman:1986uw,Kapusta:1979fh,Toimela:1984xy,coleman1988aspects,Arnold:1994ps,Arnold:1994eb,Zhai:1995ac,Chivukula:2004hw,Andersen:2011ug}.

\section{Continuum formulation}\label{Sect:ContForm}
We work in four-dimensional Euclidean space-time with coordinates $x\equiv x_\mu=(\vec x, x_4)$, with $\mu=1\ldots3$ spatial and one temporal direction $\mu=4$. Gluon fields we describe by Hermitian non-Abelian $su(N_c)$ vector fields
\begin{equation}
A_\mu(x)=gA^a_\mu(x)T_a,
\end{equation}
where  we include in analogy to the lattice formulation the gauge coupling $g$ in the definition. The index $a$ of the real-valued field components $A^a_\mu(x)$ runs over the $N_c^2-1$ gauge field components in the color direction of the Hermitian traceless generators $T_a$ of the $su(N_c)$ algebra obeying
\begin{equation}\label{algebraRules}
[T_a,T_b]=\mathrm if_{abc}T_c,\quad\mathrm{Tr}_C(T_aT_b)=\frac{1}{2}\delta_{ab},
\quad T_a=T_a^\dagger,\quad\mathrm{Tr}_C\,T_a=0,
\end{equation}
where $\mathrm{Tr}_C$ indicates the trace over the color matrices in the fundamental representation. Whereas the gauge field transforms under color transformations $\Omega(x)\in SU(N_c)$ as a connection
\begin{equation}\label{EichFreiheit}
A_\mu'(x)=\Omega(x)\,(A_\mu(x)-\mathrm i\partial_\mu)\,\Omega(x)^\dagger,
\end{equation}
the algebra-valued field strength
\begin{equation}
F_{\mu\nu}(x)=\partial_\mu A_\nu(x)-\partial_\nu A_\mu(x)+\mathrm i[A_\mu(x),A_\nu(x)]
=gF_{\mu\nu}^aT_a,
\end{equation}
transforms as a tensor
\begin{equation}
F_{\mu\nu}(x)=\Omega(x)\,F_{\mu\nu}(x)\,\Omega(x)^\dagger
\end{equation}
and  guarantees the gauge invariance of the Euclidean Yang-Mills action
\begin{equation}
S_\textrm{YM}[A]=\frac{1}{2 g^2}\int\mathrm d^4x\;\mathrm{Tr}_C(F_{\mu\nu}F_{\mu\nu})
=\frac{1}{4}\int\mathrm d^4x\;F_{\mu\nu}^aF_{\mu\nu}^a.
\end{equation}

Due to the gauge freedom~(\ref{EichFreiheit}) a perturbative vacuum $F_{\mu\nu}\equiv0$ does not necessarily mean a vanishing vector field $A_\mu$. By gauge transformations~(\ref{EichFreiheit}) a vector field $A_\mu\equiv0$ may be transformed to $A_\mu'(x)\ne0$. Moreover, gauge functions $\Omega(x)$ defined on a three-dimensional subspace of $R^4$, isomorphic to $S^3$, may have a winding number defined by the map $SU(2)\to S^3$
\begin{equation}
\Pi_3(S^3)\in\mathcal Z.
\end{equation}
The QCD-vacua are therefore characterized by an integer winding number. Transitions between neighboring winding numbers contribute to the topological charge
\begin{equation}\label{topo}
Q[A]=\frac{1}{32\pi^2}\int\mathrm d^4x\ \varepsilon_{\mu\nu\rho\sigma} 
\mathrm{Tr}_C(F_{\mu\nu}F_{\rho\sigma})\in\mathcal Z,
\end{equation}
of a field configuration. Configurations, spherical symmetric in $R^4$, with $Q=1$ and minimal action are instantons.

The gluon field mediates the interaction between $N_f$ quarks. In the path integral formulation of QCD fermions are represented by Grassmann fields. $\psi(x)$ and $\psi^\star(x)$ have independent generators of the Grassmann algebra for every $x$, flavor $f$, color $c$ and Dirac component $i$. The Euclidean Dirac matrices $\gamma_\mu$ relate the four Dirac components, {\textit e.g.}, $\overline\psi=\psi^\dagger\gamma_4$. Gauge transformations act with $\Omega(x)\in SU(N_c)$ on the color indices $c$ in the fundamental representations $\{N_c\}$ and $\{\overline{N_c}\}$
\begin{equation}\label{EichTrafoPsi}
\psi(x)'=\Omega(x)\psi(x),\quad\overline\psi(x)'=\overline\psi(x)\Omega(x)^\dagger.
\end{equation}
With $\psi(x)$ and $\overline\psi(x)$ we indicate column and row vectors with $4\times N_c$ components running over the Dirac and color components. In $\Psi(x)$ and $\overline\Psi(x)$ we include even all flavor components. For massless quarks the fermionic action is defined by
\begin{equation}\label{SfCont}
S_F[\Psi,\overline\Psi,A]=\sum_f\int\mathrm d^4x\,\overline\psi_f(x)\gamma_\mu
\left[\partial_\mu+\mathrm i\,A_\mu(x)\right]\psi_f(x)
=\int\mathrm d^4x\,\overline\Psi(x)\gamma_\mu
\left[\partial_\mu+\mathrm i\,A_\mu(x)\right]\Psi(x),
\end{equation}
which is gauge invariant by construction. We will use Euclidean Dirac matrices which are Hermitian and obey the anti-commutation relations
\begin{equation}\label{AntiGamma}
\{\gamma_\mu,\gamma_\nu\}=2\delta_{\mu\nu},\;\;\{\gamma_\mu,\gamma_5\}=0,\;\;
\gamma_5=\gamma_1\gamma_2\gamma_3\gamma_4
\quad\textrm{with}\quad\gamma_\mu^\dagger=\gamma_\mu,\;\;\gamma_5^\dagger=\gamma_5.
\end{equation}
A convenient choice of the Hermitian matrices is the Weyl- or chiral representation
\begin{eqnarray}\label{WeylGamma}
\vec\gamma=\begin{pmatrix}0&-\mathrm i\vec\sigma\\\mathrm i\vec\sigma & 0
\end{pmatrix},\qquad
\gamma_4=\begin{pmatrix}0&1\\1&0\end{pmatrix},\qquad
\gamma_5=\begin{pmatrix}1&0\\0&-1\end{pmatrix},
\end{eqnarray}
where $\vec\sigma$ are the Pauli matrices. The chiral projectors
\begin{equation}\label{ChirProj}
P_R=\frac{1+\gamma_5}{2},\quad P_L=\frac{1-\gamma_5}{2}\quad\textrm{with}\quad
P_R\,P_L=0,\;P_R^2=P_R,\;P_L^2=P_L,\;P_R\gamma_\mu=\gamma_\mu P_L
\end{equation}
map to the eigenvalues $\pm1$ of $\gamma_5$. Quark fields can therefore be decomposed into left- and right-handed components
\begin{eqnarray}\begin{aligned}
&\psi_L(x)=P_L\psi(x),\quad\psi_R(x)=P_R\psi(x),\quad 
\psi(x)=\psi_L(x)+\psi_R(x),\\
&\overline\psi_L(x)=\overline\psi(x)P_R,\quad\overline\psi_R(x)=\overline\psi(x)P_L,\quad
\overline\psi(x)=\overline\psi_L(x)+\overline\psi_R(x).
\end{aligned}\end{eqnarray}
This decomposition and the properties of the projection operators~(\ref{ChirProj}) allow to split the fermionic part~(\ref{SfCont}) of the action in the form
\begin{equation}\label{SChirDecomp}
S_F[\Psi,\overline\Psi,A]\ist{SfCont}\int\mathrm d^4x\,
\left[\overline\Psi_L(x)\,\gamma_\mu\,(\partial_\mu+\mathrm iA_\mu)\,\Psi_L(x)+
\overline\Psi_R(x)\,\gamma_\mu\,(\partial_\mu+\mathrm iA_\mu)\,\Psi_R(x)\right].
\end{equation}
This is the important ``classical'' result: the action for massless fermions decouples into two independent contributions from left- and right-handed quarks. There is no interaction term. As a result, the action of massless QCD is invariant against independent left- and right-handed transformations $U(N_f)_L\otimes U(N_f)_R$
\begin{eqnarray}\begin{aligned}
&\Psi'_L(x)=L\,\Psi_L(x),\quad\overline\Psi'(x)
=\overline\Psi_L(x)\,L^\dagger,\quad L\in U(N_f)_L,\\
&\Psi'_R(x)=R\,\Psi_R(x),\quad\overline\Psi'(x)
=\overline\Psi_R(x)\,R^\dagger,\quad R\in U(N_f)_R.
\end{aligned}\end{eqnarray}
These $U(N_f)_L\otimes U(N_f)_R$ transformations can be decomposed in vector transformations $L=R$ and axial vector transformations $L=R^\dagger$
\begin{equation}\label{UNfDecomp}
U(N_f)_L\otimes U(N_f)_R=SU(N_f)_V\otimes SU(N_f)_A\otimes U(1)_V\otimes U(1)_A.
\end{equation}
Quantum effects, the $U(1)_A$ or axial anomaly, break the $U(1)_A$ symmetry of the classical action of massless QCD, see Sect.~\ref{Sect:AnomClass}. From the remaining symmetries, the vector symmetry $U(1)_B=U(1)_V$ describes the baryon number conservation.

The mass matrix 
\begin{equation}\label{Massenmatrix}
\mathcal M=\mathrm{diag}(m_u,m_d,m_s,...,m_{N_f}).
\end{equation}
acts on the flavor indices of the fermionic fields and allows to write the mass term of the action in a compact form
\begin{equation}\label{SMass}
S_M[\Psi,\overline\Psi]=\int\mathrm d^4x\,\left[\overline\Psi_R(x)\mathcal M\Psi_L(x)+
\overline\Psi_L(x)\mathcal M^\dagger\Psi_R(x)\right].
\end{equation}
This term couples left- and right-handed fermions and therefore violates $SU(N_f)_A$, the chiral symmetry. For different quark masses the vector symmetry $SU(N_f)_V$ breaks down to 
\begin{equation}
SU(N_f)_V\otimes U(1)_V\;\to\;\prod_{f=1}^{N_f} U(1)_f
=U(1)_u\otimes U(1)_d\otimes U(1)_s\otimes\dots\otimes U(1)_{N_f}
\end{equation}
and every quark number is separately conserved. The comparison to the experiment shows that the bare quark masses $m_u$ and $m_d$ have only a few MeV, much smaller than $\Lambda_{\overline{MS}}$, and $m_s$ is of the order $\Lambda_{\overline{MS}}$. Therefore, $SU(2)_V$ is broken only slightly and ``the eightfold way'' $SU(3)_V$ more strongly. In the limit of massless u-, d- and even s-quarks the total action of QCD
\begin{equation}
S_\mathrm{QCD}[\Psi,\overline\Psi,A]
=S_\mathrm{YM}[A]+S_F[\Psi,\overline\Psi,A]+S_M[\Psi,\overline\Psi].
\end{equation}
is symmetric against chiral $SU(2)_A$ resp. $SU(3)_A$ transformations. But, in the case of spontaneous breaking of chiral symmetry the ground state of the theory does not respect this symmetry, even for $S_M[\Psi,\overline\Psi]=0$.

We should mention that for massless quarks there is only one parameter in the QCD-Lagrangian, the unit $g$ of the color charge which after renormalization turns out to be a function of the momentum transfer. There is no theoretical prediction yet explaining the values of the quark masses. Only theories beyond the standard model let us hope for an answer to this question. Quantization of the theory with the path integral
\begin{equation}
Z=\int\mathcal D[\Psi,\overline\Psi]\mathcal DA\,
\exp(-S_\mathrm{QCD}[\Psi,\overline\Psi,A]),
\end{equation}
produces infinities as long as the theory is not regularized and renormalized. A very successful perturbative regularization is dimensional regularization. Here we focus on the lattice regularization which defines QCD beyond perturbation theory.

Before discussing the lattice formulation we would like to repeat important results concerning the fermionic fields, mainly in the limit of vanishing quark masses. The fermion fields enter the QCD-action in $S_F$ and $S_M$ bilinearly. This allows to use the integration formula for Grassmann variables~\cite{Gattringer:2010zz}
\begin{equation}\label{GrassInt}
\int\mathcal D[\Psi,\overline\Psi]\,\mathrm e^{\overline\Psi\,M\,\Psi}
=\mathrm{det}\,M,
\end{equation}
where we used the matrix notation
\begin{equation}\label{MatNot}
\overline\Psi\,M\,\Psi:=\int\mathrm d^4x\,\mathrm d^4y\;\overline\Psi(x)\,M(x-y)\,\Psi(y).
\end{equation}
$\Psi(x)$ is here a column and $\overline\Psi(x)$ a row vector of Grassmann variables containing all quark components with different flavor, color and Dirac components. Moreover, if no $(x)$-dependence is indicated, like in $\Psi$ and $\overline\Psi$, the vector components run even over all coordinate values. Eq.~(\ref{GrassInt}) helps to integrate out fermions in the path integral before the integration over the gluon fields.

\section{The Axial Anomaly and the Atiyah-Singer index theorem}\label{Sect:AnomClass}
We now come back to the $U(1)_A$-anomaly, shortly mentioned after Eq.~(\ref{UNfDecomp}), in a fixed gauge background. With the decomposition~(\ref{SChirDecomp}) of the massless fermionic action and the anti-commutation relation~(\ref{AntiGamma}) one can easily see, that $S_F$ is invariant against the $L=R^\dagger\in U(1)$ axial transformations $U(1)_A$
\begin{equation}\label{gamma5Trafo}
\Psi(x)\,\to\,\Psi^\prime(x)=\exp\{\mathrm i\gamma_5\theta(x)\}\Psi(x),\quad
\overline\Psi(x)\,\to\,\overline\Psi^\prime(x)=\overline\Psi(x)\exp\{\mathrm i\gamma_5\theta(x)\}.
\end{equation}
Surprisingly, as Fujikawa~\cite{Fujikawa:1979ay} has shown, due to the infinite number of degrees of freedom in the continuum the fermionic path integral measure is not invariant under $\gamma_5$-transformations with an infinitesimal global phase $\theta$
and gives rise to an extra gluon configuration dependent phase factor
\begin{equation}\label{FujFactor}
\int\mathcal D[\Psi^\prime,\overline\Psi^\prime]=\int\mathcal D[\Psi,\overline\Psi]
\exp\{N_f\frac{2\mathrm i}{32\pi^2}\int\mathrm d^4x\,
\theta(x)\varepsilon_{\mu\nu\rho\sigma}\,\mathrm{Tr}(F_{\mu\nu}F_{\rho\sigma})\}.
\end{equation}
For constant $\theta$ and $N_f$ flavors this results in~\cite{Fujikawa:1979ay,Creutz1103.3304}
\begin{equation}\label{FujConstTheta}
\int\mathcal D[\Psi^\prime,\overline\Psi^\prime]\ist{topo}\int\mathcal D[\Psi,\overline\Psi]
\exp\{2\mathrm i\theta N_fQ[A]\}.
\end{equation}
This is the famous U$_\textrm{A}(1)$- or axial anomaly.

For the derivation~\cite{Fujikawa:1979ay,Creutz1103.3304} of Eq.~(\ref{FujFactor}) Fujikawa used the massless Dirac operator, see Eq.~(\ref{SfCont}). In our notation with Hermitian Dirac matrices~(\ref{WeylGamma})
\begin{equation}\label{masslessD}
D[A]:=\gamma_\mu(\partial_\mu+\mathrm iA_\mu),\quad D^\dagger[A]\ist{WeylGamma}-D[A],
\end{equation}
$D[A]$ is anti-Hermitian. For the normalized eigenvectors of $D[A]$ we use the Dirac notation $|\lambda\rangle$ (which may also include a degeneracy of the eigenmodes, especially there may be several zero modes $|0_i\rangle$)
\begin{equation}\label{EFD}
D[A]\;|\lambda\rangle=\lambda|\lambda\rangle,\quad
\langle\lambda|\lambda\rangle=1.
\end{equation}
Due to the anti-Hermiticity of $D[A]$ the eigenvalues $\lambda$ are purely imaginary
\begin{equation}\label{EFimag}
\langle\lambda|\;D[A]\;|\lambda\rangle\ist{EFD}\lambda,\quad
\lambda^\star=\langle\lambda|\;D^\dagger[A]\;|\lambda\rangle\ist{masslessD}
-\lambda.
\end{equation}
The anti-commutation property
\begin{equation}\label{Gamma5ASym}
\{D[A],\gamma_5\}\ist{AntiGamma}0
\end{equation}
implicates that the vectors $\gamma_5|\lambda\rangle$ are eigenvectors to the complex conjugate eigenvalues $\lambda^\star=-\lambda$
\begin{equation}\label{negEF}
|-\lambda\rangle:=\gamma_5|\lambda\rangle\quad\textrm{and}\quad
D^\dagger[A]\ist{masslessD}-D[A]\ist{Gamma5ASym}\gamma_5\;D[A]\;\gamma_5.
\end{equation}
Therefore, the eigenvalues appear in complex conjugate pairs $\lambda$ and $-\lambda$ or are zero. Restricted to the space with $\lambda=0$ we can read Eq.~(\ref{Gamma5ASym}) as commutativity of $D[A]$ and $\gamma_5$, $D[A]\gamma_5|0_i\rangle=0=\gamma_5D[A]|0_i\rangle$. In this $\lambda=0$\,-\,subspace we can diagonalize $\gamma_5$ and get zero modes of definite chirality
\begin{equation}\label{goodGamma5}
\gamma_5\;|0_j\rangle=\pm|0_j\rangle\quad\Leftrightarrow\quad
\langle0_j|\,\gamma_5\,|0_j\rangle=\pm1
\end{equation}
and their numbers $n_+$ and $n_-$ can be counted. Further we realize: Since eigenvectors to different eigenvalues are orthogonal
\begin{equation}\label{orthEF}
\langle\lambda|-\lambda\rangle\ist{negEF}
\langle\lambda|\gamma_5|\lambda\rangle=0,\quad\lambda\ne-\lambda
\end{equation}
the complex conjugate pairs do not contribute to the expectation value of the $\gamma_5$ operator. Therefore, only zero modes contribute to this expectation value
\begin{equation}\label{expG5}
\sum_\lambda\langle\lambda|\gamma_5|\lambda\rangle\ist{goodGamma5}n_+-n_-
\end{equation}
and therefore the transformation~(\ref{gamma5Trafo}) of the path integral measure~(\ref{FujConstTheta}) reads
\begin{equation}\label{FujFerm}
\int\mathcal D[\Psi^\prime,\overline\Psi^\prime]\ist{gamma5Trafo}
\int\mathcal D[\Psi,\overline\Psi]\,\mathrm{det}\,
\left(\exp\{-2\mathrm i\theta\gamma_5\}\right)
=\int\mathcal D[\Psi,\overline\Psi]\,\prod_\lambda\,
\exp\{-2\mathrm i\theta\langle\lambda|\gamma_5|\lambda\rangle\}.
\end{equation}
The minus sign in the exponent of Eq.~(\ref{FujFerm}) takes into account that the measure has to transform inverse to the fields in the integrand. Inserting Eqs.~(\ref{orthEF}) and~(\ref{goodGamma5}) we get
\begin{equation}\label{FujFerm2}
\int\mathcal D[\Psi^\prime,\overline\Psi^\prime]\ist{gamma5Trafo}
\int\mathcal D[\Psi,\overline\Psi]\,\exp\{-2\mathrm i\theta(n_+-n_-)\}.
\end{equation}
The comparison with the gluonic evaluation~(\ref{FujConstTheta}), presented in Refs.~\cite{Fujikawa:1979ay,Creutz1103.3304}, leads to the result
\begin{equation}\label{ASTheorCont}
N_f\,Q[A]=n_--n_+.
\end{equation}
It relates a fermionic property, the numbers of zero modes to a gluonic quantity, the topological charge $Q[A]$, an integer number counting the vacuum to vacuum transitions of a continuous gauge field, defined in Eq.~(\ref{topo}). The Dirac operator is a function of the gauge field and reflects this topological field property in the number of zero modes, as Atiyah and Singer proved~\cite{Atiyah:1971rm} in a more general mathematical framework. Eq.~(\ref{ASTheorCont}) results therefore from an application of the Atiyah-Singer index theorem. It allows to determine the ``analytical'' index
\begin{equation}\label{analInd}
Q_f:=n_--n_+
\end{equation}
of the Dirac operator via a property of the gauge field and vice versa.

\section{Lattice formulation}
There are many good textbooks on the lattice formulation of QCD~\cite{Creutz:1984mg,Montvay:1994cy,Smit:2002ug,DeGrand:2006zz,Gattringer:2010zz}. We do not repeat the arguments leading to various lattice formulations of the gluonic and the fermionic Lagrangian. We use standard notations for lattice spacing $a$, spatial and temporal extent $N_s$ and $N_t$ of the lattice, inverse temperature $\beta=aN_t=1/T$, with the usual choice of natural units $k_B=1$. The limit $\beta\rightarrow\infty$ corresponds to $T\rightarrow0$ and the continuum limit corresponds to $a\rightarrow0$ while keeping $aN_s$ and $aN_t$ fixed. Here, we will concentrate on questions concerning the chiral properties, which are governed by the $\gamma$-matrices, as mentioned in Sect.~\ref{Sect:ContForm}. The lattice formulation of the fermionic action suffered for a long time from the fermion doubling problem, from further poles in the propagator around momentum components $\pm\pi/a$. A well-known fermion formulation on a four-dimensional Euclidean hypercubic lattice with sites $x$ and lattice constant $a$, which removes these doubler modes, is an action originally suggested by Wilson~\cite{Wilson:1974sk}
\begin{eqnarray}\begin{aligned}\label{WilsAct}
&D_m(x,y):=m\,\delta_{x,y}+D_\mathrm{W}(x,y),\\
&D_\mathrm{W}(x,y):=\frac{4}{a}\,\delta_{x,y}-
\frac{1}{2a}\sum_{\mu=\pm1}^{\pm4}(1-\gamma_\mu)\;U_\mu(x)\;\delta_{x+\hat\mu,y}
\textrm{ with }\gamma_{-\mu}=-\gamma_\mu,\;U_{-\mu}(x)=U_\mu^\dagger(x-\hat\mu),
\end{aligned}\end{eqnarray}
where the vectors $\hat\mu$ connect nearest neighbors in $x_\mu$-direction. $U_\mu(x)\in SU(N_c)$ is the parallel transporter from $x$ to $x+\hat\mu$. It guarantees gauge invariance. The term proportional to $\gamma_\mu$, the Dirac matrix
\begin{eqnarray}\label{lattDirac}
D_L:=\gamma_\mu\nabla_\mu
:=\frac{1}{2a}\sum_{\mu=1}^4\gamma_\mu\left[U_\mu(x))\;\delta_{x+\hat\mu,y}-U_\mu^\dagger(x-\hat\mu)\;\delta_{x-\hat\mu,y}\right],
\end{eqnarray}
is the lattice version of the classical Dirac operator $D[A]$ of Eq.~(\ref{masslessD}). The remaining $a$-dependent terms in the Wilson action~(\ref{WilsAct}) are proportional to the lattice Laplacian
\begin{eqnarray}\label{lattLapl}
\Delta_L:=\sum_{\mu=1}^4\Delta_\mu
:=\sum_{\mu=1}^4\frac{1}{a^2}\left[U_\mu(x)\;\delta_{x+\hat\mu,y}-2\delta_{x,y}+U_\mu^\dagger(x-\hat\mu)\;\delta_{x-\hat\mu,y}\right]
\end{eqnarray}
and to the lattice constant $a$ and vanish therefore in the limit $a\to 0$. This momentum dependent ``$-\frac{a}{2}\Delta_L$'' term in Eq.~(\ref{WilsAct}), the Wilson term, increases effectively the mass $m$ of the doublers by $2/a$ for each momentum component around $p_\mu=\pi/a$, suppressing therefore the influence of the the doubler modes. But, like the $m$-dependent mass term, the Wilson term contains no factor $\gamma_\mu$ and therefore violates chiral symmetry; a violation which vanishes in the continuum limit only. The fermion doubling problem inhibited for a long time the treatment of dynamical fermions on the lattice. The problem was condensed in a No-go theorem by Nielsen and Ninomiya~\cite{Nielsen:1981hk}. Since the solution of the doubling problem is tightly connected to the anomaly and to chiral symmetry breaking we are now going to discuss in more detail, how  fermions can be formulated on the lattice with continuous chiral symmetry and without species doubling.

As mentioned in the paragraph before Eq.~(\ref{EichTrafoPsi}) we formulate fermions with elements of a complex Grassmann algebra for every $x$, flavor $f$, color $c$ and Dirac component $i$. With $\psi(x)$ we indicate a column vector with the $4*N_c$ components of a single flavor. In $\Psi(x)$ the $N_f$ flavor components and in the $4*N_c*N_f*V$-dimensional column $\Psi$ the $V_L$ sites of the Grassmann fields are included.

The Dirac matrix $D_L$ of Eq.~(\ref{lattDirac}) anti-commutes with $\gamma_5$
\begin{equation}\label{AnticommDg5}
\{D_L,\gamma_5\}\ist{AntiGamma}0.
\end{equation}
Therefore, it has the same classical symmetry as the continuum massless Dirac operator $D[A]$ of Eq.~(\ref{masslessD}), a symmetry not respected by quantum theory due to the infinite number of degrees of freedom in the continuum, as we discussed in Sect.~\ref{Sect:AnomClass}.

On the lattice the number of degrees of freedom is finite and therefore the fermionic measure on the lattice is obviously invariant under a global chiral transformation~(\ref{gamma5Trafo}). Since the lattice should reproduce up to order $a$ effects the same spectrum as the continuum theory there must be a mechanism in the lattice formulation breaking the U$_\textrm{A}(1)$-symmetry. A fermion formulation on the lattice with the requested property was suggested by Neuberger~\cite{Neuberger:1997fp,Neuberger:1998my,Neuberger:1998wv}. For his ``overlap'' fermions the chiral transformation~(\ref{gamma5Trafo}) is modified to~\cite{Luscher:1998pqa}
\begin{equation}\label{ChiTrafoLatt}
\Psi^\prime=\exp\{\mathrm i\theta\gamma_5(1-\frac{a}{2}D)\}\Psi,\quad
\overline\Psi^\prime=\overline\Psi\exp\{\mathrm i\theta(1-\frac{a}{2}D)\gamma_5\},
\end{equation}
where $a$ is the lattice constant and $D$ is an appropriately chosen modification of the naive lattice Dirac matrix $D_L$. Looking carefully at the requested invariance
\begin{equation}\label{ModChirInvSf}
\overline\Psi^\prime D\Psi^\prime\ist{ChiTrafoLatt}
\overline\Psi\exp\left\{\mathrm i\theta(1-\frac{a}{2}D)\gamma_5\right\}\,
D\,\exp\left\{\mathrm i\theta\gamma_5(1-\frac{a}{2}D)\right\}\Psi
\req \overline\Psi\,D\,\Psi.
\end{equation}
we realize that the invariance~(\ref{AnticommDg5}) of $D_L$ has to be relaxed to
\begin{equation}\label{GinspWils}
(1-\frac{a}{2}D)\gamma_5\,D=D\,\gamma_5(-1+\frac{a}{2}D)\quad\Leftrightarrow
\quad D\,\gamma_5+\gamma_5\,D=aD\gamma_5\,D.
\end{equation}
This invariance can be applied to every factor in the power series of the exponential in Eq.~(\ref{ChiTrafoLatt}), $[(1-\frac{a}{2}D)\gamma_5]^nD=D[\gamma_5(-1+\frac{a}{2}D)]^n$, and warrants the invariance~(\ref{ModChirInvSf}) of the Lagrangian. Eq.~(\ref{GinspWils}) is the celebrated Ginsparg--Wilson relation~\cite{Ginsparg:1981bj} which had remained unnoticed~\cite{Hasenfratz:1998ri} for a long time. The term $aD\gamma_5\,D$ in Eq.~(\ref{GinspWils}) breaks the $\gamma_5$-symmetry~(\ref{AnticommDg5}) explicitly, $D\,\gamma_5+\gamma_5\,D\ne0$. It is a term of the order $a$ which vanishes in the continuum limit. This leads to the interesting interpretation that the fermion doubling problem and the famous Nielsen-Ninomiya No-go theorem~\cite{Nielsen:1981hk} -- there are no lattice fermions without species doubling and with continuous chiral symmetry -- is a manifestation of the anomaly in the flavor singlet axial current~\cite{Karsten:1980wd}.

We want to emphasize that, as shown by L\"uscher in Ref.~\cite{Luscher:1998pqa}, for actions fulfilling the Ginsparg--Wilson relation~\cite{Ginsparg:1981bj} the transformation~(\ref{ChiTrafoLatt}) defines the symmetry~(\ref{ModChirInvSf}), which is exact at any given lattice spacing $a$. For $a\to0$ this symmetry converges to the continuum chiral symmetry.

There is another request, the Dirac matrix $D$ has to fulfill, $\gamma_5$-Hermiticity~(\ref{gamma5Herm}), which we are now going to discuss. $D$ describes the interaction between gluon fields $U$ and quark fields $\Psi$. $D$ is a functional of the gauge field
\begin{equation}\label{FunctDU}
D:=D[U].
\end{equation}
To simplify the notation further on we do not write this functional dependence explicitly. In the path integral over the Grassmann valued fermion fields $\Psi$ for given gauge configuration $[U]$ and mass matrix $\mathcal M$, see Eq.~(\ref{Massenmatrix}), we can use the bilinearity of the fermionic action in $\Psi$ and Eq.~(\ref{MatNot}) and generate the fermionic determinant $\mathrm{det}\left(D+\mathcal M\right)$, a functional of the gauge field. This determinant contains a summation over all possible closed paths of quarks moving under the influence of the gauge field. To describe the reaction of the fermions on the gauge field one tries to take the fermionic determinant as a weight factor in the probability distribution defined by the Euclidean path integral. This is only possible if the fermionic determinant is real and positive
\begin{equation}\label{RealDet}
\mathrm{det}\left(D+\mathcal M\right)\req(\mathrm{det}\left(D+\mathcal M\right))^\star=\mathrm{det}\left(D+\mathcal M\right)^\dagger.
\end{equation}
Due to $\mathcal M^\dagger\ist{Massenmatrix}\mathcal M$ the postulate of the reality of the Dirac matrix can be respected by the $\gamma_5$-Hermiticity of $D$
\begin{equation}\label{gamma5Herm}
\gamma_5\,D\,\gamma_5=D^\dagger,
\end{equation}
a request, analog to the property~(\ref{negEF}) in the continuum, and fulfilled by almost all fermionic actions. It describes that the chiral partners of fermions with opposite color charge feel the same gauge field. Attributing to pairs of quarks, like u and d-quarks, the same bare mass $m$, the product of the corresponding two fermionic determinants gives the necessary non-negative weight factor.

The conditions (\ref{GinspWils}) and (\ref{gamma5Herm}) for $D$, for a ``$\gamma_5$-Hermitian Ginsparg--Wilson--Dirac operator'', lead to important consequences which we can read from
\begin{equation}\label{NormCond}
D+D^\dagger\ist{GinspWils}a\,D\,D^\dagger,\quad
D^\dagger+D\ist{GinspWils}a\,D^\dagger\,D.
\end{equation}
From these two equation we read that $D$ and $D^\dagger$ are commuting. This is the definition for $D$ to be a normal operator. The fermionic matrix $D$ can therefore be represented by an orthonormal set of eigenvectors $|\lambda\rangle$ and their eigenvalues $\lambda$
\begin{equation}\label{DiracEq}
D=\langle\lambda|\lambda|\lambda\rangle\quad\Leftrightarrow\quad
D\,|\lambda\rangle=\lambda\,|\lambda\rangle,\quad\langle\lambda|\lambda\rangle=1.
\end{equation}
We would like to emphasize that $|\lambda\rangle$ are row vectors and $\overline\Psi_\lambda$ column vectors of complex numbers for every lattice site $x$, flavor $f$, color $c$ and Dirac component $i$. The original Grassmann variables of the fermionic fields get integrated out in the fermionic path integral~(\ref{GrassInt}). After performing this integration we can work with matrices like $D$ and their determinants.

\begin{figure}[h!]
\centering
   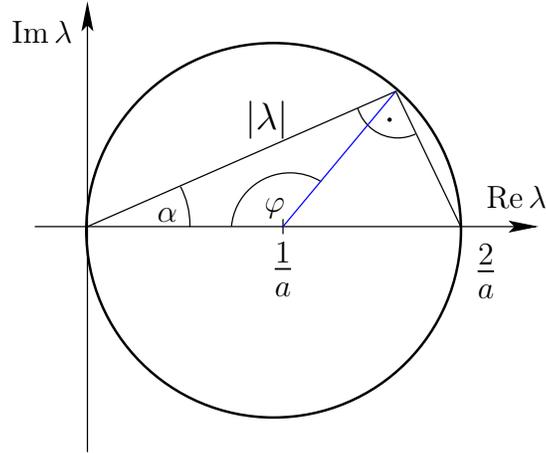
   \caption{The eigenvalues of the $\gamma_5$-Hermitian Ginsparg--Wilson--Dirac operator define the Ginsparg--Wilson circle with radius $1/a$ and $\cos\alpha=\frac{|\lambda|}{2/a}=\sin\frac{\varphi}{2}$.}
\label{fig:GWCircle}
\end{figure}

From Eq.~(\ref{gamma5Herm}) follows that $\gamma_5\,|\lambda\rangle$ is an eigenvector of $D$ to the eigenvalue $\lambda^\star$
\begin{equation}\label{ConiugPsi}
D^\dagger\ist{DiracEq}\langle\lambda|\lambda^\star|\lambda\rangle\quad
\Leftrightarrow\quad D^\dagger\;|\lambda\rangle=\lambda^\star\;|\lambda\rangle\quad
\LR{gamma5Herm}\quad D\,\gamma_5\,|\lambda\rangle\ist{gamma5Herm}\lambda^\star\,
\gamma_5\,|\lambda\rangle.
\end{equation}
Thus, the eigenvalues are either real or they appear in complex conjugate pairs $\lambda,\lambda^\star$ with eigenfunctions $|\lambda\rangle$ and $|{\lambda^\star}\rangle=\gamma_5\,|\lambda\rangle$.
From Eq.~(\ref{NormCond}) we get further interesting properties of the eigenvalues: If we insert the two eigenvalue equations~(\ref{DiracEq}) and~(\ref{ConiugPsi}) into Eq.~(\ref{NormCond}) we get
\begin{equation}\label{CondLam}
\lambda+\lambda^\star\ist{NormCond}a\,\lambda\lambda^\star.
\end{equation}
The polar representation $\lambda=|\lambda|\,\mathrm e^{\mathrm i\alpha}$ shows a nice application of Thales' theorem
\begin{equation}\label{GWCircle}
\cos\alpha=\frac{a}{2}\,|\lambda|
\end{equation}
the ``Ginsparg-Wilson circle'' of eigenvalues, see Fig.~\ref{fig:GWCircle}, where $|\lambda|$ is the hypotenuse and $2/a$ is the adjacent of $\alpha$ in a right-angled triangle.
With $1/\lambda=|\lambda|^{-1}\,\mathrm e^{-\mathrm i\alpha}$ we get simple expressions for the real and imaginary parts for $1/\lambda$ which we will use in Eq.~(\ref{separTrace}) in Sect.~\ref{Sect:ChirSymLatt}
\begin{equation}\label{rezLambda}
\frac{1}{\lambda}\ist{GWCircle}\frac{a}{2}\frac{
\mathrm e^{-\mathrm i\alpha(\lambda)}}{\cos\alpha(\lambda)}
=\frac{a}{2}-\mathrm i\,\frac{a}{2}\tan\alpha(\lambda)
=\frac{a}{2}+\frac{a}{2\mathrm i}\frac{1}{\tan\frac{\varphi(\lambda)}{2}}
\quad\textrm{since}\quad \alpha+\frac{\varphi}{2}=\frac{\pi}{2}.
\end{equation}

The eigenvectors $|\lambda\rangle$ and $|{\lambda^\star}\rangle$ are orthogonal since they belong to different eigenvalues $\lambda\ne\lambda^\star$
\begin{equation}\label{OrthoPsi}
\langle\lambda|\,\gamma_5\,|\lambda\rangle\ist{ConiugPsi}
\langle\lambda\,|{\lambda^\star}\rangle=0\quad\textrm{for}\quad
\lambda\ne\lambda^\star.
\end{equation}
This equation tells us also that the expectation value of $\gamma_5$ in the state $|\lambda\rangle$ is vanishing, if $\lambda$ is not real.

The eigenvectors to the real eigenvalues $\lambda$ are not paired. Further, they have the important property of good chirality, as can be easily seen: From the right equation in~(\ref{ConiugPsi}) we read for real $\lambda=\lambda_r\in\{0,\frac{2}{a}\}$
\begin{equation}\label{goodChir}
D\,\gamma_5\,|{\lambda_r}\rangle\ist{ConiugPsi}\lambda_r\gamma_5\,|{\lambda_r}\rangle
=\gamma_5\,D\,|{\lambda_r}\rangle,
\end{equation}
that $D$ and $\gamma_5$ are commuting in the subspace of $\lambda_r$ and therefore simultaneously diagonalizable. The eigenfunctions $|{\lambda_r}\rangle$ can be chosen with good chirality, they are ``chiral'',
\begin{equation}\label{goodChir0}
\gamma_5|{\lambda_r}\rangle=\pm|{\lambda_r}\rangle.
\end{equation}
Of special importance are the chiralities of the zero modes, of the modes with $\lambda=0$. With $n_+$ we count the number of right-handed or positive chirality modes and with $n_-$ the left-handed modes with negative chirality.

$\gamma_5$ is diagonal in the Weyl basis~(\ref{WeylGamma}) with $\mathrm{Tr}_D\,\gamma_5=0$ in each subspace of four free Diracspinors with the same momentum. Due to the invariance of the trace under basis transformations, it follows that also in the full basis of eigenstates of $D$ we get
\begin{equation}\label{vanTrace}
\sum_\lambda\langle\lambda|\,\gamma_5\,|\lambda\rangle=0.
\end{equation}
In lattice simulations one of the numbers $n_+$ and $n_-$ is always zero for usual anti-periodic boundary conditions. All zero modes have the same chirality. According to the vanishing trace~(\ref{vanTrace}) each zero mode with given chirality needs a chiral partner with opposite chirality. Such states are available only at $\lambda=\frac{2}{a}$. In the continuum limit $a\to0$ these ``doubler modes'' are sent to infinite eigenvalues.

Due to the orthogonality~(\ref{OrthoPsi}) only zero- and doubler-modes contribute to Eq.~(\ref{vanTrace}), with $\pm1$. We remove also the contributions of the doublers by multiplying the summands in Eq.~(\ref{vanTrace}) with the factor $a\lambda/2-1$ and get $(a\lambda/2-1)\langle\lambda|\,\gamma_5\,|\lambda\rangle=\langle\lambda|\,(aD/2-1)\,\gamma_5\,|\lambda\rangle$. Now only the zero modes contribute, with $-\langle\lambda|\,\gamma_5\,|\lambda\rangle$. This allows to count the difference between the zero modes of different chirality, the ``analytical'' index~(\ref{analInd})
\begin{equation}\label{AnalIndex}
Q_f:=\sum_\lambda\langle\lambda|\,
\left(\frac{a}{2}D-1\right)\,\gamma_5\,|\lambda\rangle=n_--n_+.
\end{equation}
This index has also appeared in the fermionic measure, in the axial anomaly in the continuum, in Sect.~\ref{Sect:AnomClass}.

\section{The Axial Anomaly on the lattice}\label{Sect:AnomLatt}
On the lattice the measure is invariant under the modified chiral transformation~(\ref{ChiTrafoLatt}), due to the finite number of integrations. But the result of the Grassmann integrations~(\ref{GrassInt}), the determinant of the fermionic matrix $D$ is not invariant. We get from Eq.~(\ref{ModChirInvSf})
\begin{equation}\label{TrafoDet}
\mathrm{det}D^\prime:=\mathrm{det}
\left[\exp\left\{\mathrm i\theta(1-\frac{a}{2}D)\gamma_5\right\}\right]\,
\mathrm{det} D\;\mathrm{det}\left[\exp\left\{\mathrm i\theta\gamma_5
(1-\frac{a}{2}D)\right\}\right].
\end{equation}
As we will see immediately below, the determinant is modified by the square of the determinant of the transformation matrix $T$. In the eigenbasis $|\lambda\rangle$ of $D$ one can easily see that the determinant of $T$ gets the exponential of the logarithm of the trace
\begin{equation}\label{trlogT}
\left(\mathrm{det}\,T\right)^2=\mathrm e^{2\mathrm{Tr}\ln T},\quad 
T:=\exp\left\{\mathrm i\theta(1-\frac{a}{2}D)\gamma_5\right\}
\end{equation}
With $\mathrm{Tr}\ln T=-\mathrm i\theta\mathrm{Tr}[\left(\frac{a}{2}D-1\right)\gamma_5]=-\mathrm i\theta\mathrm{Tr}[\gamma_5\left(\frac{a}{2}D-1\right)]$ we can immediately apply Eq.~(\ref{AnalIndex}) and get
\begin{equation}\label{AnomLatt}
\mathrm{det}\,D^\prime\ist{TrafoDet}\exp\{-2\mathrm i\theta N_f Q_f\}\,\mathrm{det}\,D,
\quad\textrm{with}\quad Q_f\ist{AnalIndex}n_--n_+.
\end{equation}
Up to the sign in the phase this is the same result as Fujikawa got in the continuum. The different sign reflects the fact that on the lattice we transform the fermionic matrix $D$~(\ref{TrafoDet}) which transforms inverse to the fermionic measure~(\ref{FujFerm2}). The comparison to the gluonic evaluation, which is not done here, leads again to a relation between the analytical index $Q_f$, a property of the Dirac matrix, and a gluonic property, the topological charge $Q[A]$
\begin{equation}\label{AtyiSing}
Q_f\ist{AnalIndex}n_--n_+=N_f\,Q[A].
\end{equation}
This relation between the analytical and the topological index was announced in 1963 by Michael Atiyah and Isadore Singer for elliptic differential operators on compact manifolds~\cite{atiyah1963}. They published various generalizations in a sequence of papers from 1968 to 1971~\cite{Atiyah:1971rm}. On the lattice, the theorem applies to any action that satisfies the Ginsparg-Wilson relation, including the Neuberger overlap action, however, it does not necessarily hold for non-Ginsparg-Wilson actions.

\section{Chiral condensate on the lattice}\label{Sect:ChirSymLatt}
The chiral $U(1)_A$-transformation~(\ref{ChiTrafoLatt}) acts on all fermions symmetrically. In Sect.~\ref{Sect:AnomLatt} it turned out that this symmetry is anomalous, it is broken by the fermionic integration. This anomaly has an experimental consequence, it prevents the $\eta^\prime$-meson with 957.8~MeV to be a Goldstone boson. Pions, with 135.0 and 139.6~MeV are much lighter then expected from their quark content. Also Kaons with 493.7 and 497.6~MeV and the $\eta$-meson with 547.9~MeV have a mass smaller then two thirds of the mass of baryons in the lowest octet. We can therefore expect that the chiral limit with $N_f=2$ is a very good approximation and $N_f=3$ is still good. In the approximation of massless quarks we can generalize $U(1)_A$ to $U(N_f)_A$. The additional global $SU(N_f)_A$-transformations
\begin{equation}\label{ChiTrafoLattSUN}
\Psi^\prime=\exp\{\mathrm i\theta_a\,T_a\gamma_5(1-\frac{a}{2}D)\}\Psi,\quad
\overline\Psi^\prime=\overline\Psi\exp\{\mathrm i\theta_a\,T_a(1-\frac{a}{2}D)\gamma_5\},
\end{equation}
do not lead to new anomalies since the $SU(N_f)$-generators $T_a$ are traceless. For a proof we evaluate in analogy to the discussion in Sect.~\ref{Sect:AnomLatt} the square of the determinant of the transformation matrix $T(T_a)$
\begin{equation}\label{TrafoDetSq}
\left[\mathrm{det}\,T(T_a)\right]^2:=\mathrm{det}^2
\left[\exp\left\{\mathrm i\theta_a\,T_a(1-\frac{a}{2}D)\gamma_5\right\}\right],
\end{equation}
use again Eq.~(\ref{trlogT})
\begin{equation}\label{TrafoDetSqInf}
\left[\mathrm{det}\,T(T_a)\right]^2\ist{trlogT}\exp\{-2\mathrm i\theta_a
\mathrm{Tr}\left[T_a\left(1-\frac{a}{2}D\right)\gamma_5\right]\}
\ist{algebraRules}1.
\end{equation}
and get no contribution from the determinant due the assumed flavor symmetry of the fermionic matrix $D$. To receive this result we perform in the trace $\mathrm{Tr}$, running over lattice sites $x$, color $c$, Dirac $i$ and flavor $f$ indices, the flavor trace first and use the vanishing trace of the $SU(N_f)$-generators.

Even for vanishing quark masses the chiral vector symmetry~(\ref{ChiTrafoLattSUN}) may be broken, left- and right-handed fermions may be coupled by the dynamics of QCD. To adjust the fermion formulation in analogy to the continuum to the modified chiral transformation~(\ref{ChiTrafoLattSUN}) we have to decompose the mass term~(\ref{SMass}) and the kinetic term~(\ref{SChirDecomp}) of the Lagrangian on the lattice using the Ginsparg-Wilson relation~(\ref{GinspWils}). The decomposition of the mass term seems to go as usual. For the kinetic term  we have to use a modification of the projection operators~(\ref{ChirProj}). Very helpful for this aim is the relation
\begin{equation}\label{ModGamma5}
\underbrace{\gamma_5\,(1-aD)}_{\hat\gamma_5}\;\underbrace{\gamma_5\,(1-aD)}
_{\hat\gamma_5}=1-a\{\underbrace{D+D^\dagger-aD^\dagger D}_{0}\}=1\quad
\Leftrightarrow\quad\hat\gamma_5^2=1\textrm{ with }
\hat\gamma_5:=\gamma_5\,(1-aD),
\end{equation}
which allows to define the modified projection operators~(\ref{ChirProj})
\begin{equation}\label{modProj}
\hat P_R:=\frac{1+\hat\gamma_5}{2},\quad\hat P_L:=\frac{1-\hat\gamma_5}{2}
\quad\textrm{with}\quad\hat P_R^2=\hat P_R,\quad\hat P_L^2=\hat P_L,
\quad\hat P_R\,\hat P_L=0,\quad\hat P_R+\hat P_L=1.
\end{equation}
For the decomposition of the fermionic matrix $D$ we use another way of writing the Ginsparg-Wilson relation~(\ref{GinspWils}) and get the asymmetric relations
\begin{equation}\label{asymRel}
D\,\hat P_R=(D+\underbrace{D\gamma_5-a\,D\gamma_5D}_{\ist{GinspWils}-\gamma_5D})/2=
P_L\,D,\quad D\,\hat P_L=(D-\underbrace{D\gamma_5-a\,D\gamma_5D}
_{\ist{GinspWils}-\gamma_5D})/2=P_R\,D.
\end{equation}
We apply therefore different projectors for $\Psi$ and $\overline\Psi$
\begin{equation}\label{psiDecomp}
\overline\Psi_R:=\overline\Psi\,P_L,\quad\overline\Psi_L:=\overline\Psi\,P_R,\quad
\Psi_R:=\hat P_R\,\Psi,\quad\Psi_L:=\hat P_L\,\Psi
\end{equation}
and arrive at a decomposition of the fermionic matrix in analogy to the continuum
\begin{equation}\label{LattDecompD}
\overline\Psi\,D\,\Psi=\overline\Psi_R\,D\,\Psi_R+\overline\Psi_L\,D\,\Psi_L.
\end{equation}
The asymmetric definition~(\ref{psiDecomp}) of the projected wave functions has consequences for the mass term~(\ref{SMass}). If we define this term on the lattice with the projected components
\begin{equation}\label{LattMassTerm}
\overline\Psi_L\mathcal M\Psi_R+\overline\Psi_R\mathcal M\Psi_L\ist{psiDecomp}
\overline\Psi\mathcal M(\underbrace{P_R\hat P_R}_{P_R(1-\frac{a}{2}D)}
+\underbrace{P_L\hat P_L}_{P_L(1-\frac{a}{2}D)})\Psi
=\overline\Psi\mathcal M(1-\frac{a}{2}D)\Psi
\end{equation}
we get an additional term leading via the fermionic matrix $D$ to an additional coupling of neighboring lattice sites. We combine this mass term with the kinetic term and get the total fermionic action of Ginsparg-Wilson fermions
\begin{equation}\label{TotalSGW}
S_{GW}:=\overline\Psi\,\left(1-\frac{a\mathcal M}{2}\right)\,D\,\Psi
+\overline\Psi\,\mathcal M\,\Psi=\overline\Psi\,D_\mathcal M\,\Psi
\quad\textrm{mit}\quad D_{\mathcal M}:=D+\mathcal M\left(1-\frac{a}{2}\,D\right).
\end{equation}
As mentioned before, $\Psi$ is a column vector and $\overline\Psi$ a row vector containing Grassmann variables for every lattice site $x$, flavor $f$, color $c$ and Dirac component $i$. For the further calculations it is important that the mass matrix $\mathcal M$ is block diagonal in the flavor quantum number $f$. Below, we need especially the submatrix $D^u$ for the lightest quark flavor $u$ with mass $m$ and the submatrix
\begin{equation}\label{subDm}
D_m:\ist{TotalSGW}D^u+m\left(1-\frac{a}{2}\,D^u\right).
\end{equation}
of $D_{\mathcal M}$.

If chiral symmetry would be intact left- and right-handed quarks could be transformed independently and there would be no coupling between left- and right-handed quarks, a coupling as it appears in the mass term~(\ref{LattMassTerm}). If the invariance is lost, we call the symmetry broken and define an order parameter for the breaking of chiral symmetry, the ``quark condensate'' $\Sigma$. For its definition~(\ref{quarkKondens}) we use the scalar expectation value
\begin{equation}\label{quarkKondmV}
\Sigma(a,m,V_L):\ist{LattMassTerm}-\frac{1}{a^4V_L}
\left\langle\bar u\,\left(1-\frac{a}{2}\,D^u\right)\,u\right\rangle
\end{equation}
of the bilinear $\bar u(x)u(x)$ with a form indicated by the mass term~(\ref{LattMassTerm}) for the lightest Grassmann valued quark field $u(x)$ in the infinite volume $V=a^4V_L$, vanishing $u$-quark mass $m$ and continuum $a\to0$ limit. The fermionic bilinear $\bar u\,\left(1-\frac{a}{2}\,D^u\right)\,u$ includes besides sums over Dirac indices $i$ and color indices $c$ also a sum over lattice sites $x$, therefore we divide in Eq.~(\ref{quarkKondmV}) by the physical volume $V$. The brackets $\langle\rangle$ in Eq.~(\ref{quarkKondmV}) indicate the gluonic and fermionic path integrals. According to Eq.~(\ref{GrassInt}) the Grassmann integration over the fermionic bilinear leads to a subdeterminant, where one row and one column, corresponding to the components of $u(x)$ and $\bar u(x)$, of the fermionic matrix are removed. Consequently, the subdeterminants for flavors $f\ne u$ are unaffected and only the subdeterminant of the $u$-quarks has to be treated in detail. From linear algebra we know that a matrix element of the inverse matrix is just such a subdeterminant divided by the determinant. Taking all this into account we get
\begin{equation}\label{quarkKondmV1}
\Sigma(a,m,V_L)\ist{quarkKondmV}\frac{1}{a^4V_L}\left\langle\mathrm{Tr}
\left[\left(1-\frac{a}{2}\,D^u\right)\,D_m^{-1}\right]\right\rangle_G.
\end{equation}
The sign change from Eq.~(\ref{quarkKondmV}) to Eq.~(\ref{quarkKondmV1}) originates in the sign of the exponent of the fermionic Boltzmann factor $\exp\{-S_{GW}\}$. The matrices $D^u$ and $D_m$ and the trace $\mathrm{Tr}$ in Eq.~(\ref{quarkKondmV1}) extend only over the $u$-quark degrees of freedom and the full determinant of $D_{\mathcal M}$ is included in the gluonic path integral
\begin{equation}\label{gluonInt}
\langle O\rangle_G:=\frac{1}{Z}\int\mathcal DU
\mathrm e^{-S_g[U]}\,O[U]\,\mathrm{det}(D_{\mathcal M}[U]).
\end{equation}
 The study of the dependence of $\Sigma(a,m,V_L)$ on the volume $V$, the quark mass $m$ and the lattice constant $a$ will be discussed in Sect.~\ref{sec:numevlqcd}. Here we describe the analytic evaluation of the path integral~(\ref{quarkKondmV}) for given $a$ in the $V\to\infty$ and  $m\to0$, the ``chiral'' limit. 

In the $m\to0$ limit of Eq.~(\ref{quarkKondmV}) we find a nice interpretation of the factor $1-\frac{a}{2}D$ characteristic for Ginsparg-Wilson fermions, which appeared in the chiral transformation~(\ref{ChiTrafoLattSUN}) and in the mass term~(\ref{LattMassTerm})
\begin{equation}\label{mEqZero}
\lim_{m\to0}\mathrm{Tr}\left[\left(1-\frac{a}{2}\,D^u\right)\,D_m^{-1}\right]
\ist{subDm}\mathrm{Tr}\left(\frac{1}{D^u}-\frac{a}{2}\right)\ist{DiracEq}
\sum_\lambda\left(\frac{1}{\lambda}-\frac{a}{2}\right)\ist{rezLambda}
\mathrm i\sum_\lambda\mathrm{Im}\frac{1}{\lambda}.
\end{equation}
We recognize that in the $m\to0$ limit the factor $1-\frac{a}{2}D^u$ removes the real part of the eigenvalues $1/\lambda$ of $1/D^u$. Further we realize that for the evaluation of the trace in Eqs.~(\ref{quarkKondmV1}) and~(\ref{mEqZero}) we should treat the eigenvectors for $\lambda=0$ separately. For the ``operator'' in expression~(\ref{quarkKondmV1}) we get
\begin{equation}\begin{aligned}\label{separTrace}
\mathrm{Tr}&\left[\left(1-\frac{a}{2}\,D^u\right)\,D_m^{-1}\right]\ist{subDm}
\mathrm{Tr}\,\frac{1-\frac{a}{2}\,D^u}{m\left(1-\frac{a}{2}\,D^u\right)+D^u
}=\mathrm{Tr}\,\frac{1}{m+\frac{D^u}{1-\frac{a}{2}\,D^u}}=\\
&\ist{DiracEq}\sum_{\lambda=0}\frac{1}{m}+\sum_{\lambda(\ne\lambda^\star)}
\frac{1}{m+\frac{1}{\frac{1}{\lambda}-\frac{a}{2}}}
\ist{rezLambda}\frac{n_++n_-}{m}+\sum_{\lambda(\ne\lambda^\star)}
\frac{1}{2}\left[\frac{1}{m+\mathrm i\frac{2}{a}\tan\frac{\varphi_\lambda}{2}}+
\frac{1}{m-\mathrm i\frac{2}{a}\tan\frac{\varphi_\lambda}{2}}\right],
\end{aligned}\end{equation}
where $n_+$ and $n_-$ are the numbers of right-handed and left-handed zero modes and $\varphi_\lambda$ is the angle $\varphi$ of Fig.~\ref{fig:GWCircle} related to the complex eigenvalue $\lambda$ on the Ginsparg-Wilson circle with positive or negative imaginary part. Extending in Eq.~(\ref{separTrace}) the summation over all non-real eigenvalues $\lambda\ne\lambda^\star$ we want to indicate also that the doublers with $\lambda=2/a$ do not contribute to the trace. Due to the factor $1/V$ in Eq.~(\ref{quarkKondmV1}) also the zero modes do not contribute in the $V\to\infty$ limit since their number $n_++n_-$ does not grow faster than $\sqrt{V}$.

For continuous
\begin{equation}\label{varphiApprox}
y:=\frac{2}{a}\tan\frac{\varphi_\lambda}{2}\approx
\frac{\varphi_\lambda}{a}\approx-\mathrm i\lambda=\pm|\lambda|
\quad\textrm{for}\quad\varphi\approx0
\end{equation}
the expression in the square bracket of Eq.~(\ref{separTrace}) approaches in the $m\to0$ limit a $\delta$-function at $\lambda=0$
\begin{equation}\label{DeltaApprox}
\lim_{m\to0}\frac{1}{2}\left[\frac{1}{m+\mathrm iy}+\frac{1}{m-\mathrm iy}
\right]=\pi\delta(y)\ist{varphiApprox}
\pi\delta(\frac{\varphi_\lambda}{a}),\quad\int\frac{m}{m^2+y^2}\,\mathrm dy
=\int\frac{\mathrm d\frac{y}{m}}{1+\frac{y^2}{m^2}}
=\left.\mathrm{atan}\frac{y}{m}\right|_{-\infty}^{+\infty}=\pi.
\end{equation}
Approaching the sum over $\lambda$ in Eq.~(\ref{separTrace}) by an integral over a density function $\rho_\lambda$ we can formally perform an integral over the delta-function. Due to the distribution of eigenvalues on the Ginsparg-Wilson circle we expect for given $m$ a density function $\rho_\lambda(\varphi,m)$ to be approximately independent of $a$ and the number of eigenvalues to increase with $V_L$. Including the gluonic average and the remaining factor $a^{-4}$ of Eq.~(\ref{quarkKondmV1}) we define therefore
\begin{equation}\label{Udens}
\frac{1}{a^4}\left\langle\frac{1}{V_L}\sum_{|\lambda|<\Lambda}\right\rangle_G\;\to\;
\frac{1}{a^4}\int_{-a\Lambda}^{a\Lambda}\mathrm d\varphi\,\rho_\lambda(\varphi,m).
\end{equation}
Combining Eq.~(\ref{DeltaApprox}) with Eq.~(\ref{Udens}) we define the chiral condensate
\begin{equation}\begin{aligned}\label{quarkKondens}
\Sigma:=&\lim_{a\to0}\lim_{m\to0}\lim_{V_L\to\infty}\Sigma(a,m,V_L)\ist{quarkKondmV}
\lim_{a\to0}\lim_{m\to0}\lim_{V_L\to\infty}\frac{\pi}{a^4}\int\mathrm d\varphi\,
\rho_\lambda(\varphi,m)\,\delta(\frac{\varphi}{a})=\\
=&\lim_{a\to0}\lim_{m\to0}\lim_{V_L\to\infty}\frac{\pi}{a^3}\rho_\lambda(0,m):=\pi\rho(0).
\end{aligned}\end{equation}
This result is the famous Banks-Casher relation~\cite{Banks:1979yr}.

To break a symmetry spontaneously requires an infinite system, the limit of the number $V_L$ of lattice points going to infinity. Like in other cases of spontaneous symmetry breaking the $V\to\infty$, the ``thermodynamic'' limit, cannot be interchanged with the removal of the explicit symmetry breaking term, in our case the $m\to0$ ``chiral'' limit. The results of numerical calculations should be independent of the order of the other limits.

In lattice calculations the Banks-Casher relation is an optimal tool to determine the value of the chiral condensate from the density of near-zero modes of the Dirac operator. If this density is non-vanishing then the chiral symmetry is broken. It is interesting to know the value of the condensate for various gluonic coupling constants, quark masses, baryonic densities and temperatures.

For completeness we should mention that the chiral condensate has not only been obtained using the Banks-Casher relation but also from the Gell-Mann-Oakes-Renner relation and using the Wilson flow. Finally, the eigenvalue distribution of the Dirac operator can also be obtained through the application of (chiral) Random Matrix Theory to QCD, where the detailed dependence of the partition function and Dirac operator eigenvalue correlation functions on finite lattice spacing $a$ or chemical potential $\mu$ are computed. For an excellent review and recent developments see~\cite{Akemann:2016keq} and references therein. 

\section{Numerical evidence for chiral symmetry breaking}\label{sec:numevlqcd}
The chiral condensate $\Sigma$ varies under chiral transformations and serves therefore as a convenient order parameter of chiral symmetry breaking. Chiral symmetry breaking is a non-perturbative phenomenon and can be analyzed on the lattice and in chiral perturbation theory ($\chi$PT). As proposed in Ref.~\cite{0812.3638} an efficient method to determine the value of $\Sigma$ in lattice calculations is based on the Banks-Casher relation~(\ref{quarkKondens}), see Ref.~\cite{Banks:1979yr}. It uses the condensation of low modes of the Dirac operator near the origin. In any numerical determination of a quantity the results of a computation need an extrapolation to certain limits, in this case the limits indicated in Eq.~(\ref{quarkKondens}). The functional dependence on the extrapolation parameters can be derived within the extended framework of $\chi$PT~\cite{Sharpe:2006ia,Necco:2013sxa} and the Gell-Mann-Oakes-Renner (GMOR) relation~\cite{GellMann:1968rz}
\begin{equation}\label{GMOR}
\lim_{m\to0}\frac{M_\pi^2F_\pi^2}{2m}=\Sigma.
\end{equation}
According to this relation the derivative of $M_\pi^2F_\pi^2/2$ with respect to the quark mass $m$ in the chiral limit must be equal to this condensation rate. $M_\pi$ and $F_\pi$ are here the mass and the decay rate of the Nambu-Goldstone bosons. We use the convention $F_\pi\approx90$~MeV and $N_f$=2.

The first step in numerical determination of $\Sigma$ based on the Banks-Casher relation~(\ref{quarkKondens}) is the calculation~(\ref{DiracEq}) of the lowest eigenvalues $\lambda$ of the Dirac operator $D$ for the lightest quark flavor $u$ with mass $m$
\begin{equation}\label{DefDm}
D\,u_\lambda=\lambda u_\lambda
\end{equation}
For $D=D_L$ of Eq.~(\ref{lattDirac}) the eigenvalues are purely imaginary, for the $\gamma_5$-Hermitian Dirac matrices~(\ref{gamma5Herm}) they appear in complex conjugate pairs or are real, and for fermions obeying the Ginsparg--Wilson relation~(\ref{NormCond}), $D\gamma_5+\gamma_5D=aD\gamma_5D$, they are distributed on the Ginsparg-Wilson circle. The computer codes usually deliver the absolute values of $\lambda$ in units of the lattice constant $a$. Small values of $\lambda$ are directly related to the central angle $\varphi$ of the Ginsparg-Wilson circle in Fig.~\ref{fig:GWCircle}, $\mathrm i\varphi_\lambda\approx a\lambda$, see Eq.~(\ref{varphiApprox}). The density $\rho_\lambda(\varphi,m)$ of eigenvalues for given $\varphi$ and $m$ is approximately independent of the lattice constant $a$ and increasing with number $V_L$ of lattice sites. The value of the chiral condensate follows from the spectral density near zero via the Banks-Casher relation~(\ref{quarkKondens}).

Computationally more efficient than the determination of $\rho_\lambda(\varphi,m)$ is the integral over the spectral density, the gluonic average over the number of modes in the interval $[-\Lambda,\Lambda]$ 
\begin{equation}\label{nuLambda}
\nu(\Lambda,m):=\left\langle\sum_{|\lambda|<\Lambda}\right\rangle_G\;\zu{Udens}\;
V_L\int_{-a\Lambda}^{a\Lambda}\mathrm d\varphi\,\rho_\lambda(\varphi,m),
\end{equation}
It is quite understandable that in the $\Lambda$-region around zero $\nu(\Lambda,m)$ is proportional to $V_L$. For the free case one can derive $\nu(\Lambda,m)\propto V_L\Lambda^4$, from Eq.~(\ref{nuLambda}) one is getting $\rho_\lambda(\varphi)\propto\varphi^3$ and therefore a vanishing condensate~(\ref{quarkKondens}). In contrast to this result the numerical calculations show that in the chirally broken phase $\nu$ of Eq.~(\ref{nuLambda}) grows proportional to $\Lambda$
\begin{equation}\label{ModNum}
\nu(\Lambda,m)\ist{quarkKondens}V_L\left(2a\Lambda\right)
\left(\frac{a^3}{\pi}\Sigma\right)+\dots=\frac{2V}{\pi}\,\Lambda\,\Sigma+\dots,
\end{equation}
where the symmetry of the integration region $[-a\Lambda,a\Lambda]$ explains the factor $2$ .

Recently, the mode number $\nu(\Lambda,m)$ for two light quark flavors has been computed for the tree-level Symanzik improved gluon action and the Wilson twisted mass fermion action by Refs.~\cite{Cichy:2011an,Cichy:2013gja} and for the standard Wilson gluonic action and the non-perturbatively O(a)-improved Wilson fermion action~\cite{Luscher:1996ug} by Refs.~\cite{1406.4987,Engel:2014eea,1511.08786}. As an example for the determination of the chiral condensate on the lattice we follow the discussion in Ref.~\cite{1511.08786}.

The mode number $\nu(\Lambda,m)$ of Eq.~(\ref{nuLambda}) is equal to the average number of eigenvalues $\alpha$ of the massive Hermitian Dirac operator $D_m^\dagger D_m=(D_\mathrm{W}^\dagger+m)(D_\mathrm{W}+m)$ according Eq.~(\ref{WilsAct}) with $\alpha\le\Lambda^2 + m^2$. Expression~(\ref{nuLambda}) gives the so called ``bare chiral condensate'' at a given lattice constant $a$ only. As QCD is a renormalizable theory the ``physical'' value of the chiral condensate depends on the regularization scheme and can be converted to other schemes by the appropriate renormalization factors $Z$. As proven in Ref.~\cite{0812.3638} the rate of condensation is renormalizable and unambiguously defined after renormalization of the bare action parameters. Therefore, the following results, see also Refs.~\cite{DelDebbio:2005qa,0812.3638,1406.4987,Engel:2014eea,1511.08786}, can be computed using the (improved) Wilson formulation of lattice QCD even though the latter violates chiral symmetry at energies on the order of the inverse lattice spacing. We conclude that the mode number $\nu$ is a renormalization group invariant quantity
\begin{equation}
\nu_\mathrm{R}(\Lambda_\mathrm{R},m_\mathrm{R})=\nu(\Lambda,m),\quad
\Lambda_\mathrm{R}=Z_m\Lambda,\quad m_R=Z_mm.
\end{equation}
The renormalized chiral condensate $\Sigma_\mathrm{R}$ can be deduced from the discretized derivative of Eq.~(\ref{ModNum}), by the ``effective spectral density''
\begin{equation}\label{EffSpecDen}
\tilde\rho_\mathrm{R}(\Lambda_\mathrm{R},m_\mathrm{R})=\frac{\pi}{2V}\frac{\nu_\mathrm{R}(\Lambda_\mathrm{R1},m_\mathrm{R})-\nu_\mathrm{R}(\Lambda_\mathrm{R1},m_\mathrm{R})}{\Lambda_\mathrm{R2}-\Lambda_\mathrm{R1}},\quad\Lambda_\mathrm{R}=\frac{\Lambda_\mathrm{R2}+\Lambda_\mathrm{R1}}{2}.
\end{equation}

\begin{figure}[h!]
\centering
   \includegraphics[scale=0.66]{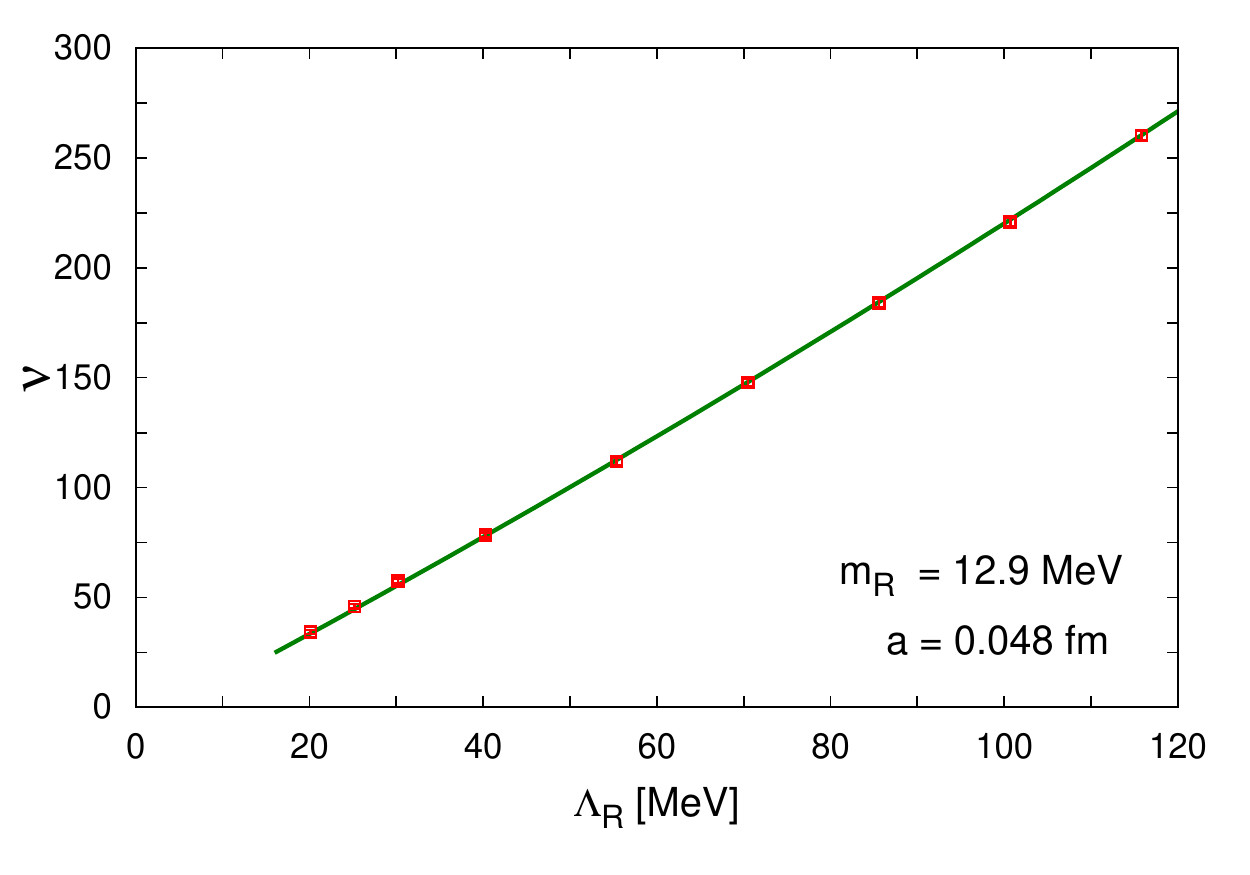}
   \includegraphics[scale=0.66]{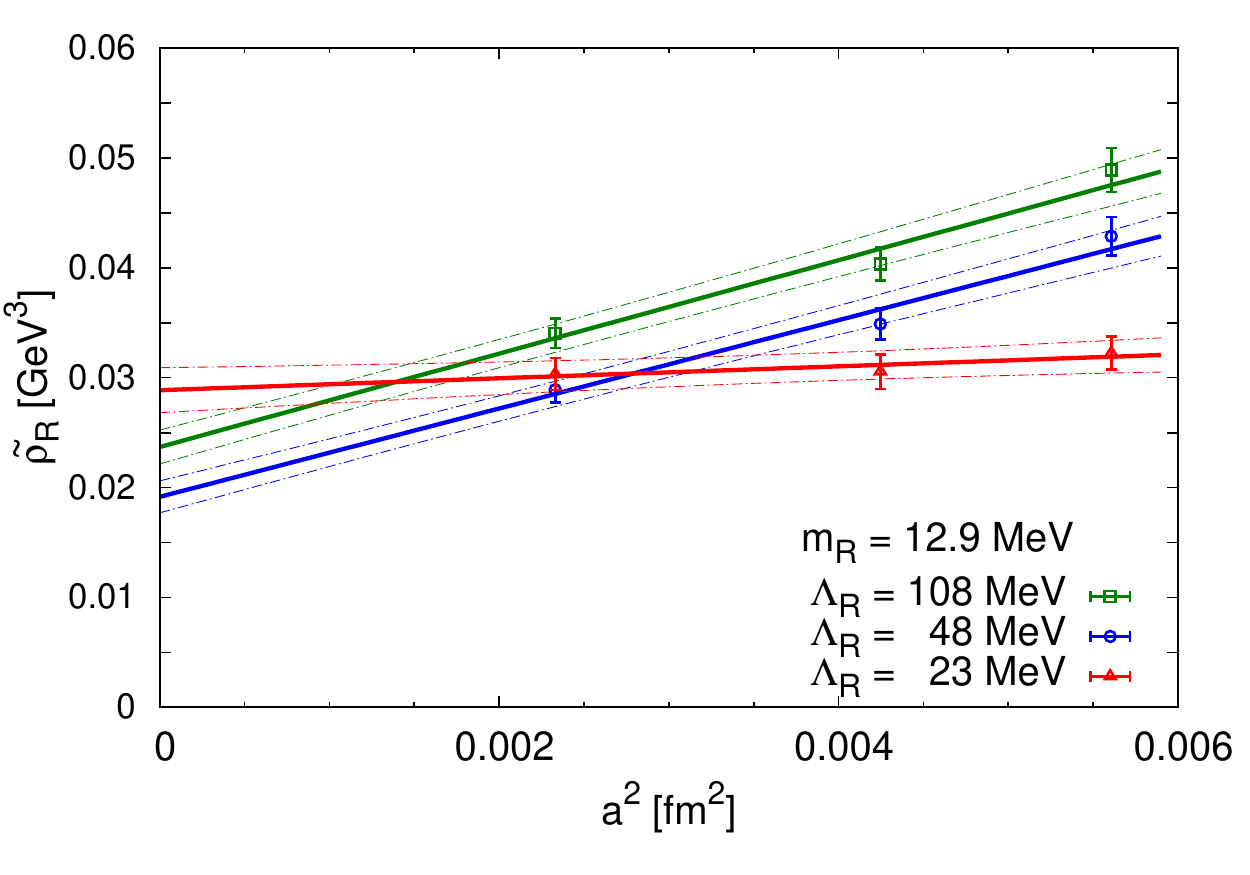}
   \caption{Left: The mode number $\nu_\mathrm{R}(\Lambda_\mathrm{R},m_\mathrm{R})$ as a function of $\Lambda_\mathrm{R}$ with a quadratic fit to the data $\nu_\mathrm{R}=-9.0(13)+2.07(7)\Lambda_\mathrm{R}/\mathrm{MeV}+0.0022(4)(\Lambda_\mathrm{R}/\mathrm{MeV})^2$. $m_R/m_s\approx0.14$. Right: Linear fits in $a^2$ to the effective spectral density $\tilde\rho_\mathrm{R}(\Lambda_\mathrm{R},m_\mathrm{R})$ for three values of $\Lambda_\mathrm{R}$ and fixed $m_\mathrm{R}$. Courtesy of~\cite{Engel:2014eea}.}
\label{fig:ModeNum}
\end{figure}

The left diagram in Fig.~\ref{fig:ModeNum} depicts the result of the determination of the mode number for nine values of $\Lambda_\mathrm{R}$ for given lattice constant $a$ and two light fermions of mass $m_\mathrm{R}=12.9$~MeV ($m_R/m_s^{\overline{\textrm{MS}}}\approx0.126$ according to Eq.~(7.9) of Ref.~\cite{Fritzsch:2012wq}) on a $64^3$x$128$ lattice. The renormalized values are given in the $\overline{\textrm{MS}}$-scheme at the renormalization scale $\mu=2$~GeV. As the quadratic fit to the data $\nu_\mathrm{R}=-9.0(13)+2.07(7)\Lambda_\mathrm{R}/\mathrm{MeV}+0.0022(4)(\Lambda_\mathrm{R}/\mathrm{MeV})^2$ shows, the main contribution to $\nu_\mathrm{R}$ is linear in $\Lambda_\mathrm{R}$ while the constant and quadratic term are in the investigated region of the order of 10\%. This agrees with the expectation from the Banks--Casher relation.

\begin{figure}[h!]
   \includegraphics[scale=0.72]{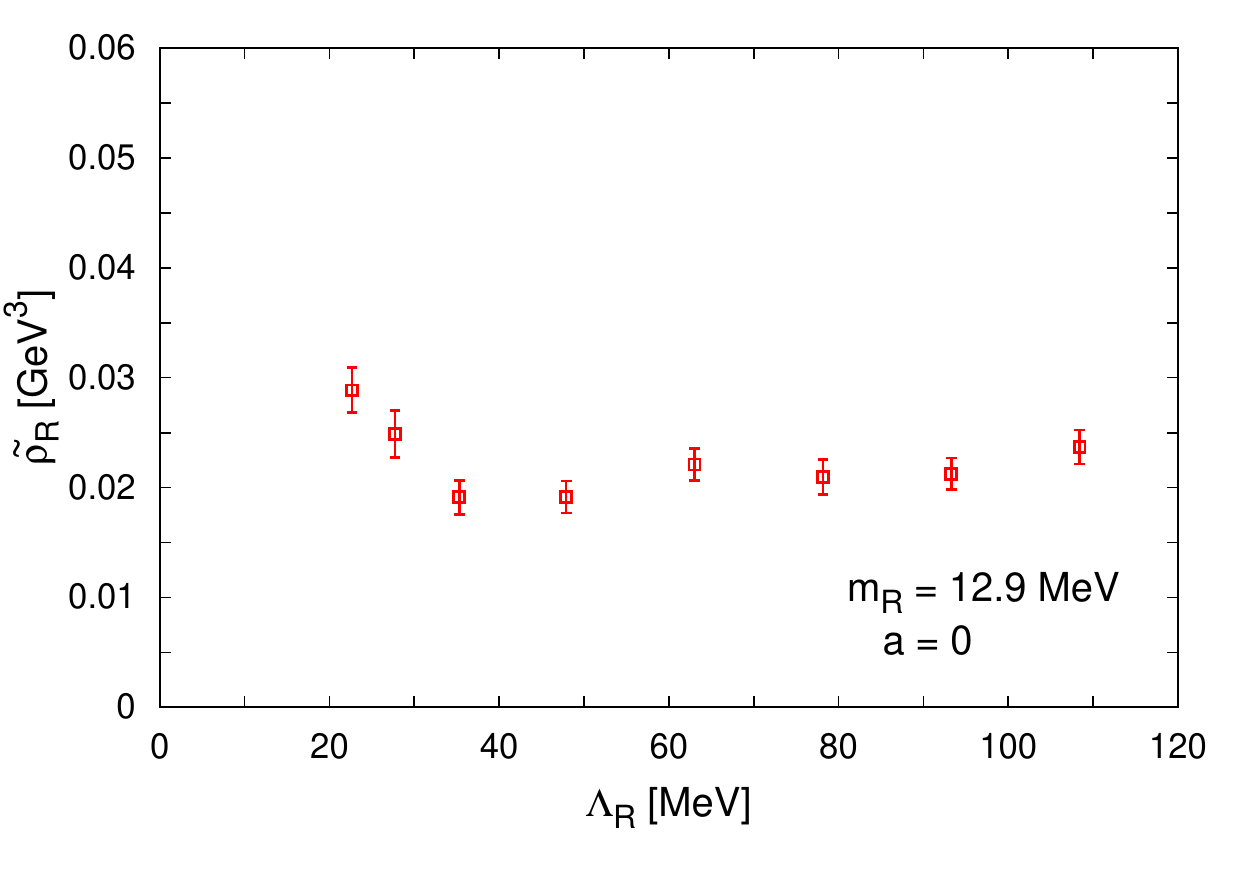}
   \includegraphics[scale=0.72]{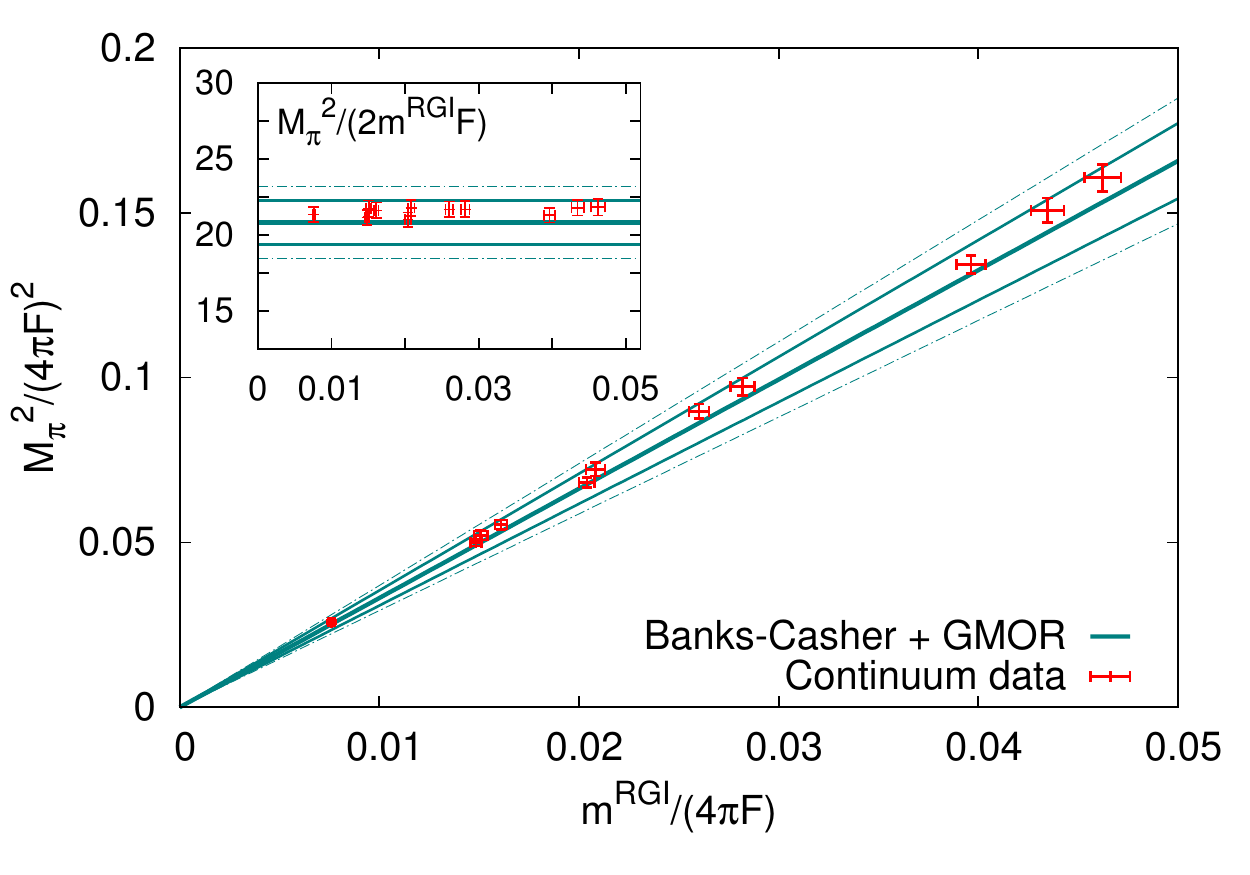} 
   \caption{Left: Continuum limit of $\tilde\rho_\mathrm{R}(\Lambda_\mathrm{R},m_\mathrm{R})$ for the same quark mass $m_\mathrm{R}=12.9$~MeV ($m_R/m_s\approx0.14$). Right: The pion mass squared versus the RGI quark mass, both normalized to $4 \pi F$ which is roughly $1$~GeV. The ratio $(M_\pi^2/2m F)^{1/3}$ is extrapolated to the continuum. The central line is the GMOR contribution to the pion mass squared computed by taking the direct measure of the condensate from the spectral density. The upper and lower solid lines show the statistical error and the dotted-dashed ones the total error, the systematic being added in quadrature. Courtesy of~\cite{Engel:2014eea}.}
\label{fig:ContLim}
\end{figure}

From the eight couples of consecutive values of $\Lambda_\mathrm{R}$ eight values of the effective spectral density $\tilde\rho_\mathrm{R}$ are determined with Eq.~(\ref{EffSpecDen}) for the three considered values of $m_\mathrm{R}$ and three lattice spacings $a$ in Ref.~\cite{Engel:2014eea}. This allows at fixed quark masses $m_\mathrm{R}$ and averaged $\Lambda_\mathrm{R}$-values a fit with a linear function in $a^2$, as the O(a)-improved fermionic action predicts, see the right diagram of Fig.~\ref{fig:ModeNum}. The result of an extrapolation to the continuum limit for the eight averaged $\Lambda_\mathrm{R}$-values and the quark mass $m_\mathrm{R}=12.9$~MeV and of the GMOR shown in Fig.~\ref{fig:ContLim}. It should be emphasized that without any assumption about the existence of spontaneous symmetry breaking the left diagram of Fig.~\ref{fig:ContLim} shows that for sufficiently small quark masses $m_\mathrm{R}$ the density of near-zero modes is finite and to a large extend independent of the boundary $\Lambda_\mathrm{R}$ of the averaging region. The right diagram of Fig.~\ref{fig:ContLim} shows that the spectral density of the Dirac operator in the continuum is non-zero at the origin and that its value agrees with the slope of $M_\pi^2 F_\pi^2/2$ with respect to the quark mass when both are extrapolated to the chiral limit. If expanded in $m$, $M_\pi^2$ is dominated by the leading (GMOR) term proportional to the chiral condensate. The ratio $M_\pi^2/2m$ is nearly constant within errors up to quark masses that are about one order of magnitude larger than in Nature.

\begin{figure}[h!]
\centering
   \includegraphics[scale=0.85]{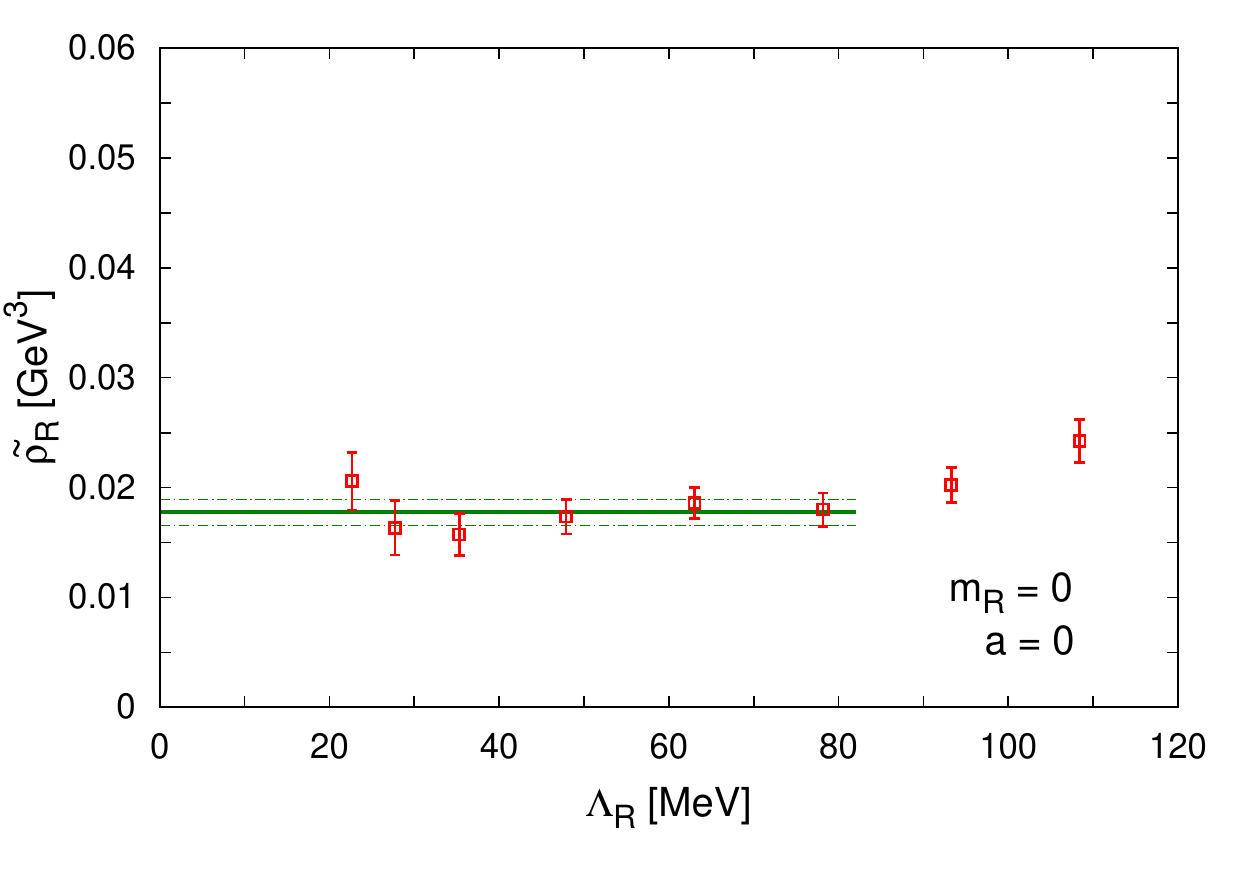}
   \caption{Effective spectral density $\rho_\mathrm{R}$ versus the cutoff $\Lambda_\mathrm{R}$ in the continuum and chiral limits. The constant line gives the value for the chiral condensate. Courtesy of~\cite{Engel:2014eea}.}
\label{fig:ChirContLim}
\end{figure}

The chiral limit, on the other hand, requires an assumption on how $\rho_\mathrm{R}(\Lambda_\mathrm{R},m_\mathrm{R})$ behaves for $m_\mathrm{R}\to0$. A corresponding framework is given by chiral perturbation theory ($\chi$PT). $\chi$PT is an effective theory based on the spontaneous breaking of chiral symmetry and a soft explicit breaking by quark-mass terms. $\chi$PT predicts in next to leading order $\rho_\mathrm{R}$ to be $\Lambda_\mathrm{R}$-independent~\cite{Engel:2014eea}. By different fits inspired from $\chi$PT Ref.~\cite{Engel:2014eea}  extrapolates $\rho_\mathrm{R}(\Lambda_\mathrm{R},m_\mathrm{R})$ to the chiral and the continuum limit, see Fig.~\ref{fig:ChirContLim}, and finds for the chiral condensate values of $\Sigma^{1/3}=$261(6)~MeV, 253(9)~MeV and by fitting the data in all three directions $\Lambda_\mathrm{R},m_\mathrm{R}$ and $a$ at the same time 259(6)~MeV. 
These results are in good agreement with earlier investigations in quenched lattice QCD with exact chiral symmetry \cite{Chiu:2003iw}, where a chiral condensate of $\Sigma^{1/3}=$250(3)~MeV in the $\overline{\textrm{MS}}$-scheme at the renormalization scale 2~GeV was determined. 

After this detailed discussion of a lattice determination of the chiral condensate $\Sigma$ we would like to present as a result of the recent compilation~\cite{Aoki:2016frl} the estimates of the $N_f=2$ and $N_f=2+$1 condensates in the $\overline{\textrm{MS}}$-scheme at the renormalization scale 2~GeV
\begin{equation}\begin{aligned}\label{AokiComp}
&N_f=2:\quad&&\Sigma^{1/3}=266(10)~\textrm{MeV}\quad&&\textrm{Refs.~\cite{Engel:2014eea,Cichy:2013gja,Baron:2009wt}},\\
&N_f=2+1:\quad&&\Sigma^{1/3}=274(3)~\textrm{MeV}\quad&&\textrm{Refs.~\cite{Blum:2014tka,Bazavov:2010yq,Borsanyi:2012zv,Durr:2013goa}}.
\end{aligned}\end{equation}
The errors include both statistical and systematic uncertainties.

More recently, the authors of~\cite{Cossu:2016eqs} compute the chiral condensate in 2+1-flavor QCD through the spectrum of low-lying eigenmodes of Dirac operator. The number of eigenvalues of the Dirac operator is evaluated using a stochastic method with an eigenvalue filtering technique on the background gauge configurations generated by lattice QCD simulations including the effects of dynamical up, down and strange quarks described by the M\"obius domain-wall fermion formulation. The spectrum shape and its dependence on the sea quark masses calculated in numerical simulations are consistent with the expectation from one-loop chiral perturbation theory. After taking the chiral and continuum limits using the data at three lattice spacings ranging $0.080-0.045$ fm, they obtain $\Sigma^{1/3}=270(4.9)$MeV, with the error combining statistical and various sources of systematic errors. Finite volume effects are confirmed to be under control by a direct comparison of the results from two different volumes at the lightest available sea quarks corresponding to $230$ MeV pions. JLQCD and TWQCD Collaborations~\cite{Fukaya:2009fh,Fukaya:2010na} find slightly lower values for the chiral condensate using lattice QCD, chiral Random Matrix Theory and chiral perturbation theory with $N_f=2, 2+1$ and $3$.

In the last few years, the Yang-Mills gradient flow was shown to be an attractive tool for non-perturbative studies of non-Abelian gauge theories. In view of its renormalization properties~\cite{Luscher:2010iy,Luscher:2011bx}, and since its application in lattice gauge theory is technically straightforward, the gradient or Wilson flow allows the dynamics of non-Abelian gauge theories to be probed in many interesting ways. The flow can be used for accurate scale setting, for example, and it provides an understanding of how exactly the topological (instanton) sectors emerge in the continuum limit of lattice QCD~\cite{Luscher:2010iy}. Moreover, observables at positive flow time are natural quantities to consider for non-perturbative renormalization and step scaling~\cite{Luscher:1991wu,Fodor:2012td,Fodor:2012qh,Fritzsch:2013je}. Matter fields may or may not be included in the flow equations. A fairly trivial extension of the flow to the quark fields in QCD is achieved, by leaving the flow equation for the gauge field unchanged, while the evolution of the quark fields as a function of the flow time is determined by a gauge-covariant heat equation. Ref.~\cite{Luscher:2013cpa} gives an excellent introduction and overview of the gradient flow in QCD and illustrates two applications of the extended flow with respect to chiral symmetry, one being a new strategy for the calculation of the axial-current renormalization constant in lattice QCD and the other a computation of the chiral condensate essentially through the evaluation of the expectation value of the scalar quark density at positive flow time. In both cases, the method is technically attractive, the chiral condensate, for example, is easily obtained with high precision, because no additive renormalization is required. 

\section{Towards a mechanism of chiral symmetry breaking}\label{sec:concl}
Quantum chromodynamics (QCD) at low energies is dominated by the non-perturbative phenomena of quark confinement and (spontaneous) chiral symmetry breaking ($\chi$SB). A rigorous treatment of them is only possible in the lattice regularization and the interplay between $\chi$SB and confinement as well as the chiral and deconfinement phase transitions at finite temperature and density are of continuous interests \cite{Polyakov:1978vu,'tHooft:1977hy,Casher:1979vw,Banks:1979yr,Hatta:2003ga,Mocsy:2003qw,Marhauser:2008fz,Braun:2007bx,Braun:2009gm}. 
 
The origin of $\chi$SB may be described as an analog to magnetization. Its strength is measured by the fermion (chiral) condensate in Eq.~(\ref{quarkKondmV}), which is an order parameter for $\chi$SB. It is a vacuum condensate of bilinear expressions involving the quarks in the QCD vacuum. The trivial,  perturbative vacuum is the field configuration with lowest action. The path integral formulation of quantum field theories underlines the importance of entropy. Due to the entropy contribution the QCD vacuum is non-trivial, dominated by quantum fluctuations, leading to the prominent non-perturbative phenomena, confinement and chiral symmetry breaking.

The entropy may be enhanced due to additional minima of the action and due to symmetries. There is an infinite number of vacua of QCD, characterized by an integer winding number. Instantons and anti-instantons, transitions between vacua with neighboring winding numbers, are relative minima of the action. They inspire the instanton picture of the QCD vacuum~\cite{Belavin:1975fg,Actor:1979in,'tHooft:1976fv,Bernard:1979qt}. On the lattice we identify another symmetry of the action, center symmetry. It is the basis for the appearance of percolating quantized magnetic flux, center vortices, forming surfaces in four-dimensional space~\cite{'tHooft:1977hy,Vinciarelli:1978kp,Yoneya:1978dt,Cornwall:1979hz,Mack:1978rq,Nielsen:1979xu}. Independent piercings of Wilson loops by vortices lead to a confining string tension with a strength depending on their density which is determined by the coupling constant~\cite{DelDebbio:1996mh,Langfeld:1997jx,DelDebbio:1997ke,Kovacs:1998xm,Engelhardt:1999wr,Bertle:2002mm,Engelhardt:2003wm}. This density defines a length scale for the quantum theory and breaks the scale symmetry explicitly, it is anomalous. Vortices may have a topological non-trivial color structure. Abelian projection of colorful vortices defines Abelian magnetic monopoles on vortices, supporting the monopole picture of confinement. Intersections, writhing points and color structure lead to lumps of topological charge and relate the vortex and the instanton picture of the QCD-vacuum~\cite{deForcrand:1999ms,Alexandrou:1999vx,Reinhardt:2000ck,Engelhardt:2002qs,Bornyakov:2007fz,hollwieser:2008tq,Hollwieser:2011uj,Schweigler:2012ae,Hollwieser:2012kb,Hollwieser:2013xja,Hoellwieser:2014isa,Hollwieser:2014mxa,Hollwieser:2014lxa,Greensite:2014gra,Hollwieser:2015koa,Trewartha:2015nna,Hollwieser:2015qea,Altarawneh:2015bya,Altarawneh:2016ped,Trewartha:2017ive,Faber:2017alm,Biddle:2018dtc}. The Atiyah-Singer index theorem connects the total topological charge of gluonic field configurations with the number of zero modes, which induce the $U(1)_A$-anomaly. Interacting lumps of topological charge lead to low-lying Dirac modes which via the Banks-Casher relation determine the strength of SB$\chi$S. Hence field configurations with lumps of topological charge increase the density of low-lying Dirac eigenmodes with pronounced local chiral properties producing a finite chiral condensate. 

This is a ``kinematical'' scenario for SB$\chi$S. Let us try to conjecture a ``dynamical'' picture: The low momentum modes of quark fields change chirality, when they enter a combination of parallel color electric and magnetic fields, present in regions of non-vanishing topological charge density. Such fields force slow color charges into spiraling paths changing their momentum and conserving their spin. Fast moving charges are less influenced by such field combinations. This could explain the importance of low-lying Dirac modes for SB$\chi$S and clarify why Goldstone bosons do not survive the removal of low-lying Dirac modes and heavy hadrons with increasing removal tend to increase their masses. 

There are many unsolved interesting problems concerning the vacuum structure of QCD, confinement and chiral symmetry breaking. Some of the most interesting questions for future work to generate progress in
this field are as follows:

\begin{itemize}
\item Do chirally polarized low-energy modes condense? What is the physical origin of the band width $\Lambda_\mathrm{ch}$ of condensing modes?
\item Does the result that fermionic zero modes and chirality are localized on structures with fractal dimension $D=2-3$ favor the vortex/domain-wall nature of the localization?
\item Which kind of effective quark-gluon interactions are generated by dynamical $\chi$SB? Will this include a scalar confining force?
\item Why do Goldstone bosons not survive the removal of low-lying Dirac modes?
\item What is the relative contribution of the various interacting topological objects to the Dirac operator's density of modes around zero virtuality?
\item Do low-momentum modes change chirality in regions of non-vanishing topological charge density with electric {\bf and} magnetic fields present and thus dynamically break chiral symmetry?
\item Can one construct an explicit quantum state responsible for a dissipation-free flow of an electric current along an external magnetic field (chiral magnetic effect)?
\end{itemize}
An answer to these questions may hold the key to understand infrared QCD and the related phenomena, most prominently, confinement and dynamical $\chi$SB.

\section{Conclusions}

We reviewed the most prominent non-perturbative features of QCD, in particular the various aspects of chiral symmetry breaking ($\chi$SB): i) the dynamical $\chi$SB that leads to the pions being light
pseudo-Goldstone bosons; ii) the anomaly, which eliminates the flavor-singlet axial U(1) symmetry
and prevents the $\eta'$-meson with a mass of order $\Lambda$QCD to be a Goldstone boson; iii) the explicit symmetry breaking from the quark masses, responsible for the pseudo-scalar mesons not being exactly
massless. We presented numerical evidence for $\chi$SB and discussed its restoration at finite temperature and density. The understanding of the mechanisms goes well beyond perturbation theory and
a rigorous treatment of them is presently only possible in the lattice regularization. For a more detailed review see~\cite{Faber:2017alm}.

\section*{Acknowledgments}
We thank Dmitry Antonov, Gerhard Ecker, Michael Engelhardt, Jeff Greensite, Urs M. Heller, Derek Leinweber and {\v S}tefan Olejn\'{\i}k for helpful discussions. This research was supported by an Erwin Schr\"odinger Fellowship of the Austrian Science Fund FWF (``Fonds zur F\"orderung der wissenschaftlichen Forschung'') under Contract No. J3425-N27 (R.H.).

\bibliographystyle{utphys}
\bibliography{chsb}

\end{document}